\documentclass[conference]{IEEEtran}

\let\labelindent\relax %
\usepackage[utf8]{inputenc}
\usepackage{tcolorbox}
\usepackage{xspace}
\usepackage{color}
\usepackage{soul}
\usepackage[nohyperlinks, nolist]{acronym}
\usepackage{amsmath}
\usepackage{wasysym}
\usepackage[hidelinks]{hyperref}
\usepackage{cleveref}
\usepackage{tikz}
\usetikzlibrary{calc, arrows}
\usepackage{algpseudocode}
\usepackage{enumitem}
\usepackage{amsthm}
\usepackage{calc}
\usepackage{booktabs}
\usepackage{array}
\usepackage{multirow}
\usepackage{pifont}
\usepackage{mathtools}
\usepackage[normalem]{ulem}
\usepackage{tikz}
\usetikzlibrary{shapes.geometric}

\newcommand{\sysname}{\textsl{POSE}\xspace}

\hypersetup{hidelinks}

\tcbuselibrary{breakable}

\newtcolorbox[auto counter]{boxProgram}[2][]{colback=white,colbacktitle=white,coltitle=black,colframe=black,arc=0pt, auto outer arc, boxrule=0.5pt, fontupper={\footnotesize}, title=\textbf{Program \thetcbcounter}: #2, floatplacement=t, #1}

\newtheoremstyle{allTheorems}%
{}%
{}%
{\itshape}%
{}%
{\bfseries}%
{:}%
{ }%
{}%

\theoremstyle{allTheorems}
\newtheorem{claim}{Claim}

\newcommand{\txt}[1]{\mathit{#1}}
\newcommand{\txtMsg}[1]{\mathtt{#1}}

\newcommand{\define}{:=}

\newcommand{\pk}{\txt{pk}}
\newcommand{\sk}{\txt{sk}}
\newcommand{\GenSig}{\txt{GenSig}}
\newcommand{\GenPK}{\txt{GenPK}}
\newcommand{\Enc}{\txt{Enc}}
\newcommand{\Dec}{\txt{Dec}}
\newcommand{\Sign}{\txt{Sign}}
\newcommand{\Verify}{\txt{Verify}}
\newcommand{\Hash}{H}

\newcommand{\poolsize}{n}

\newcommand{\ledger}{\mathsf{BC}}
\newcommand{\ledgerHeaders}{\mathsf{BCH}}
\newcommand{\enclave}{\mathcal{E}}
\newcommand{\tee}{T}	%
\newcommand{\teeSet}{\mathcal{T}}
\newcommand{\manager}{M}

\newcommand{\user}{U}

\newcommand{\TEE}{\txt{TEE}}
\newcommand{\install}{\txt{install}}
\newcommand{\resume}{\txt{resume}}
\newcommand{\prog}{\txt{prog}}
\newcommand{\VerifyQuote}{\txt{VerifyQuote}}

\newcommand{\latestBock}{\txt{now}}
\newcommand{\nbConfBlocks}{\gamma}
\newcommand{\blockTime}{\tau}
\newcommand{\confTime}{\Delta}

\newcommand{\tees}{\txt{registered}}	%

\newcommand{\conCreator}{\txt{creator}}
\newcommand{\conCodeHash}{\txt{codeHash}}
\newcommand{\conPool}{\txt{pool}}
\newcommand{\conExecChalMsg}{\txt{c1Msg}}
\newcommand{\conWatchChalMsg}{\txt{c2Msg}}
\newcommand{\conExecChalBlock}{\txt{c1Time}}
\newcommand{\conWatchChalBlock}{\txt{c2Time}}
\newcommand{\conExecChalRes}{\txt{c1Res}}
\newcommand{\conWatchChalRes}{\txt{c2Res}}
\newcommand{\conPayoutLevel}{\txt{payouts}}
\newcommand{\conBalance}{\txt{balance}}

\newcommand{\received}{\txt{Rec}}
\newcommand{\checkpoint}{p}
\newcommand{\conId}{\txt{id}}
\newcommand{\state}{\txt{state}}
\newcommand{\code}{\txt{code}}
\newcommand{\move}{\txt{move}}
\newcommand{\blockHash}{\txt{bh}}
\newcommand{\flag}{\txt{flag}}
\newcommand{\openResponses}{\teeSet_{\txt{wait}}}
\newcommand{\key}{\txt{key}}
\newcommand{\pool}{\teeSet}
\newcommand{\coins}{\txt{amount}}

\newcommand{\timeout}{\delta}
\newcommand{\timeOffchainExecution}{\timeout^1_{\txt{off}}}
\newcommand{\timeOffchainPropagation}{\timeout^2_{\txt{off}}}
\newcommand{\timeOnchainExecution}{\timeout^1_{\txt{on}}}
\newcommand{\timeOnchainPropagation}{\timeout^2_{\txt{on}}}
\newcommand{\timeOnchainCreation}{\timeOnchainExecution}
\newcommand{\timeOffchainCreation}{\timeOffchainExecution}
\newcommand{\timeOnchainCreationPropagation}{\timeOnchainPropagation}

\newcommand{\protCreationChallenge}{\txt{CreationChallenge}}
\newcommand{\protWatchdogChallenge}{\txt{WatchdogChallenge}}
\newcommand{\protWatchdogCreationChallenge}{\txt{WatchdogCreationChallenge}}
\newcommand{\protExecutiveChallenge}{\txt{ExecutiveChallenge}}
\newcommand{\Validate}{\txt{Validate}}

\newcommand{\initContract}{\txt{initContract}}

\newcommand{\conNextState}{\txt{nextState}}
\newcommand{\conGetState}{\txt{getState}}
\newcommand{\conUpdateState}{\txt{update}}

\newcommand{\pre}{\txt{pre}}
\newcommand{\conf}{\txt{conf}}
\newcommand{\res}{\txt{res}}

\newcommand{\msgDeposit}{\txtMsg{deposit}}
\newcommand{\msgWithdraw}{\txtMsg{withdraw}}
\newcommand{\msgRegister}{\txtMsg{register}}

\newcommand{\msgCreate}{\txtMsg{create}}
\newcommand{\msgInit}{\txtMsg{init}}
\newcommand{\msgExecute}{\txtMsg{execute}}
\newcommand{\msgBad}{\txtMsg{bad}}
\newcommand{\msgUpdate}{\txtMsg{update}}
\newcommand{\msgConfirm}{\txtMsg{confirm}}
\newcommand{\msgOk}{\txtMsg{ok}}

\newcommand{\msgFinalize}{\txtMsg{finalize}}

\begin{acronym}
	\acro{tee}[TEE]{Trusted Execution Environment}
	\acro{mpc}[MPC]{Multiparty Computation}
\end{acronym}
	
\begin{document}

\title{POSE: \textbf{P}ractical \textbf{O}ff-chain \textbf{S}mart Contract \textbf{E}xecution \\ (Full Version)}

\date{}

\author{\IEEEauthorblockN{Tommaso Frassetto\IEEEauthorrefmark{1}, Patrick Jauernig\IEEEauthorrefmark{1}, David Koisser\IEEEauthorrefmark{1}, David Kretzler\IEEEauthorrefmark{2}, \\ Benjamin Schlosser\IEEEauthorrefmark{2}, Sebastian Faust\IEEEauthorrefmark{2} and Ahmad-Reza Sadeghi\IEEEauthorrefmark{1}}
	\IEEEauthorblockA{Technical University of Darmstadt, Germany}
\IEEEauthorblockA{\IEEEauthorrefmark{1}first.last@trust.tu-darmstadt.de}\IEEEauthorblockA{\IEEEauthorrefmark{2}first.last@tu-darmstadt.de}}

\maketitle

\begin{abstract}
	Smart contracts enable users to execute payments depending on complex program logic. 
	Ethereum is the most notable example of a blockchain that supports smart contracts leveraged for countless applications including games, auctions and financial products. 
	Unfortunately, the traditional method of running contract code \emph{on-chain} is very expensive, for instance, on the Ethereum platform, fees have dramatically increased, rendering the system unsuitable for complex applications.
	A prominent solution to address this problem is to execute code \emph{off-chain} and only use the blockchain as a trust anchor. 
	While there has been significant progress in developing off-chain systems over the last years, current off-chain solutions suffer from various drawbacks including costly blockchain interactions, lack of data privacy, huge capital costs from locked collateral, or supporting only a restricted set of applications.

	In this paper, we present \sysname---a practical off-chain protocol for smart contracts that addresses the aforementioned shortcomings of existing solutions.
	\sysname leverages a pool of Trusted Execution Environments (TEEs) to execute the computation efficiently and to swiftly recover from accidental or malicious failures.
	We show that \sysname provides strong security guarantees even if a large subset of parties is corrupted.
	We evaluate our proof-of-concept implementation with respect to its efficiency and effectiveness.
\end{abstract}

\section{Introduction}
\label{sec:intro}
More than a decade ago, Bitcoin~\cite{bitcoin} introduced the idea of a decentralized cryptocurrency, marking the advent of the blockchain era.
Since then, blockchain technologies have rapidly evolved and a plethora of innovations emerged with the aim to replace centralized platform providers by distributed systems.
One particularly important application of blockchains concerns so-called \emph{smart contracts}, complex transactions executing payments that depend on programs deployed to the blockchain.
The first and most popular blockchain platform that supported complex smart contracts is Ethereum~\cite{ethereum}.
However, Ethereum still falls short of the decentralized ``world computer'' that was envisioned by the community~\cite{ethereum-world-computer}.
For example, contracts are replicated among a large group of miners, thereby severely limiting scalability and leading to high costs.
As a result, most contracts used in practice in the Ethereum ecosystem are very simple: 80\% of popular contracts consist of less than 211 instructions, and almost half of the most active contracts are simple token managers~\cite{oliva2020exploratory}.
More recently proposed computing platforms in permissionless decentralized settings (e.g., \cite{Cardano,dfinity}) suffer from similar scalability limitations.
\\

In recent years, numerous solutions have been proposed to address these shortcomings of blockchains, one of the most promising being so-called \emph{off-chain execution} systems.
These protocols move the majority of transactions off-chain, thereby minimizing the costly interactions with the blockchain.
\newcommand*\squared[1]{%
	\tikz[baseline=(char.base)]{\node[regular polygon, regular polygon sides=4,draw,minimum size=1.5em,inner sep=0pt] (char) {#1};}}
A large body of work has explored various types of off-chain solutions including most prominently state-channels~\cite{Sprites,PerunStateChannels,L4}, Plasma~\cite{Plasma,khalil2018commit} and Rollups~\cite{arbitrumRollup,optimisitcRollup}, which are actively investigated by the Ethereum research community. Other schemes use execution agents that need to agree with each other~\cite{ace,bitcontracts}, rely on incentive mechanisms~\cite{Arbitrum,truebit}, or leverage \acp{tee}~\cite{ekiden,fastkitten}.
A core challenge that arises while designing off-chain execution protocols is to handle the possibility of parties who stop responding, either maliciously or accidentally.
Without countermeasures, this may cause the contract execution to stop unexpectedly, which violates the \emph{liveness} property.
Despite major progress towards achieving liveness in a off-chain setting, current solutions come with at least one of these limitations: \squared{i} participating parties need to lock large amounts of collateral;  \squared{ii} costly blockchain interactions are required at every step of the process or at regular intervals; and finally \squared{iii}
the set of participants and the lifetime need to be known beforehand, which limits the set of applications supported by the system.
Additionally, existing solutions often \squared{iv} do not support keeping the contract state confidential, which is required, e.g., for eBay-style proxy auctions~\cite{proxy-bid} and games such as poker.
We refer the reader to Table~\ref{tab:rw} for an overview on related work and to Section~\ref{sec:relatedwork} for a detailed discussion.

Addressing all of these limitations in one solution while guaranteeing liveness is highly challenging.
Currently, there are two ways to address the risks of unresponsive parties.
The first approach is to require collateral, i.e., parties have to block large amounts of money, which is used to disincentivize malicious behavior and to compensate parties in case of premature termination (cf.~\squared{i}).
Since the amount of collateral depends on the number of participants and the amount of money in the contract, both must be fixed for the whole lifetime of the contract.
To ensure payout of the collateral, the lifetime of the contract must be fixed as well (cf.~\squared{iii}).
The second approach is to store contract state on the blockchain to enable other parties to resume execution.
However, this is both expensive and leads to long waiting times due to frequent synchronization with the blockchain (cf. \squared{ii}).
Further, if the contract state needs to be confidential, and hence, is not publicly verifiable, verifying the correctness of the contract execution is harder (cf. \squared{iv}).
Realizing a system tackling all these challenges in a holistic way could pave the way towards the envisioned ``world computer''.
We will further elaborate on the specific challenges in \Cref{sec:design}.

\noindent \textbf{Our goals and contributions:} We present \sysname, a novel off-chain execution framework for smart contracts in permissionless blockchains that overcomes these challenges, while achieving correctness and strong liveness guarantees.
In \sysname, each smart contract runs on its own subset of \acp{tee} randomly selected from all \acp{tee} registered to the network.
One of the selected \acp{tee} is responsible for the execution of a smart contract.

However, as the system hosting the executing \ac{tee} may be malicious (e.g., the \ac{tee} could simply be powered off during contract execution), our protocol faces the challenge of dealing with malicious operator tampering, withholding and replaying messages to/from the \ac{tee}.
Hence, the \ac{tee} sends state updates to the other selected \acp{tee}, such that they can replace the executing \ac{tee} if required.
This makes \sysname the first off-chain execution protocol with strong liveness guarantees.
In particular, liveness is guaranteed as long as at least one \ac{tee} in the execution pool is responsive.
Due to this liveness guarantee, there is no inherent need for a large collateral in \sysname (cf. \squared{i}).
The state remains confidential, which allows \sysname to have private state (cf. \squared{iv}).
Furthermore, \sysname allows participants to change their stake in the contract at any time.
Thus, \sysname supports contracts without an a-priori fixed lifetime and enables the set of participants to be dynamic (cf. \squared{iii}).
Above all, \sysname executes smart contracts quickly and efficiently without any blockchain interactions in the optimistic case (cf. \squared{ii}).

This enables the execution of highly complex smart contracts and supports emerging applications to be run on the blockchain, such as federated machine learning.
Thus, \sysname improves the state of the art significantly in terms of security guarantees and smart contract features.
To summarize, we list our main contributions below:
\begin{itemize}
	\item We introduce \sysname, a fast and efficient off-chain smart contract execution protocol.
	It provides strong guarantees without relying on blockchain interactions during optimistic execution, and does not require large collaterals. Moreover, it supports contracts with an arbitrary contract lifetime and a dynamic set of users. An additional unique feature of \sysname is that it allows for confidential state execution.
	\item We provide a security analysis in a strong adversarial model. We consider an adversary which may deviate arbitrarily from the protocol description. We show that \sysname achieves correctness and state privacy as well as strong liveness guarantees under static corruption, even in a network with a large share of corrupted parties.
	\item To illustrate the feasibility of our scheme, we implement a prototype of \sysname using ARM TrustZone as the \ac{tee} and evaluated it on practical smart contracts, including one that can merge models for federated machine learning in 238ms per aggregation.
\end{itemize}

\section{Adversary Model}
\label{sec:adversary-model}

The goal of \sysname is to allow a set of users to run a complex smart contract on a number of TEE-enabled systems.
Note, that \sysname is TEE-agnostic and can be instantiated on any TEE architecture adhering to our assumptions, similar to, e.g., FastKitten~\cite{fastkitten}.
In order to model the behavior and the capabilities of every participant of the system, we make the following assumptions:

\begin{enumerate}[start=1, label=\textbf{A\arabic*:}, ref={A\arabic*}, leftmargin=0pt, labelsep=\widthof{~}, itemindent=\widthof{A1:~}]

	\item We assume the TEE to protect the enclave program, in line with other TEE-assisted blockchain proposals~\cite{zhang2016town, fastkitten, ekiden, bentov2019tesseract, zhang2018paralysis, lind2018teechain}. Specifically:

	\begin{enumerate}[start=1, label=\textbf{\theenumi.\arabic*:}, ref={\theenumi.\arabic*}, leftmargin=0pt, labelsep=\widthof{~}, itemindent=\widthof{A1.1:~}]

		\item \label{itm:TEE-i-c} We assume the TEE to provide integrity and confidentiality guarantees.
		This means that the TEE ensures that the enclave program runs correctly, is not leaking any data, and is not tampering with other enclaves.
		While our proof of concept is based on TrustZone, our design does not depend on any specific TEE.
		In practice, the security of a TEE is not always flawless, especially regarding information leaks. However, plenty of mitigations exist for the respective commercial \acp{tee}; hence, we consider the problem of information leakage from any specific TEE, as well as TEE-specific vulnerabilities in security services, orthogonal to the scope of this paper.
		We discuss some mitigations to side-channel attacks to TrustZone,
		as well as the possible grave consequences of a compromised or leaking TEE for the executed smart contract, in~\Cref{sec:architectural-security}.

		\item \label{itm:TEE-no-exploit} We further assume the adversaries to be unable to exploit memory corruption vulnerabilities in the enclave program. This could be ensured using a number of different approaches, e.g., by using memory-safe languages, by deploying a run-time defense like CFI~\cite{abadi2005cfi}, or by proving the correctness of the enclave program using formal methods. The existence of these defenses can be proven through remote attestation (cf. \ref{itm:TEE-attestation}).
		
	\end{enumerate}
	
	\item
	We assume the TEE to provide a good source of randomness to all its enclaves and to have access to a relative clock according to the GlobalPlatform TEE specification~\cite{globaltee}.

	\item \label{itm:TEE-attestation} We assume the TEE to support \emph{secure remote attestation}, i.e., to be able to provide unforgeable cryptographic proof that a specific program is running inside of a genuine, authentic enclave. Further, we assume the attestation primitive to allow differentiation of two enclaves running the same code under the same data.
	Note that today's industrial TEEs support remote attestation~\cite{samsung-attestation, qualcomm-attestation, google-attestation, johnson2016intel, sev2020strengthening}.

	\item \label{itm:TEE-operators} We assume the TEE operators, i.e., the persons or organizations owning the TEE-enabled machines, to have full control over those machines, including root access and control over the network. The operators can, for instance, provide wrong data to an enclave, delay the transmission of messages to it, or drop messages completely. The operators can also completely disconnect an enclave from the network or (equivalently) power off the machine containing it. However, as stated in \ref{itm:TEE-i-c}, the operators cannot leak data from any enclave or influence its computation in any way besides by sending (potentially malicious) messages to it through the official software interfaces.

	\item \label{itm:static-corruption} We assume static corruption by the adversary.
	More precisely, a fixed fraction of all operators is corrupted while an arbitrary number of users can be malicious (including the case where they all are).
	We model each of the malicious parties as \emph{byzantine adversaries}, i.e., they can behave in arbitrary ways.
	
	\item \label{itm:blockchain} We assume the blockchain used by the parties to satisfy the following standard security properties:
	common prefix (ignoring the last $\nbConfBlocks$ blocks, honest miners have an identical chain prefix), chain quality (blockchain of honest miner contains significant fraction of blocks created by honest miners), and chain growth (new blocks are added continuously).
	These properties imply that valid transactions are included in one of the next $\alpha$ blocks and that no valid blockchain fork of length at least $\nbConfBlocks$ can grow with the same block creation rate as the main chain.
	We deem protection against network attacks (e.g., network partition attacks), which violate these standard properties, orthogonal to our work.

\end{enumerate}

\section{Design}
\label{sec:design}
\sysname is a novel off-chain protocol for highly efficient smart contract execution, while providing strong correctness, privacy, and liveness guarantees.
To achieve this, \sysname leverages the integrity and confidentiality guarantees of TEEs to speed up contract execution and make significantly more complex contracts practical\footnote{%
We design \sysname without depending on any specific TEE implementation. 
In \Cref{sec:architectural-security}, we discuss the implications of using ARM TrustZone to realize our scheme.}.
This is in contrast to executing contracts on-chain, where computation and verification is distributed over many parties during the mining process.
\sysname supports contracts with arbitrary lifetime and number of users, which includes complex applications like the well-known CryptoKitties~\cite{cryptoKitties}.
We elaborate more on interaction between contracts in Appendix~\ref{app: supported contracts}.
Our protocol involves users, operators and a single on-chain smart contract.
\emph{Users} aim to interact with smart contracts by providing inputs and obtaining outputs in return.
\emph{Operators} own and manage the TEE-enabled systems and contribute computing power to the \sysname network by creating protected execution units, called \emph{enclaves}, using their TEEs.
These enclaves perform the actual state transitions triggered by users.
A simple on-chain smart contract, which we call \emph{manager}, is used to manage the off-chain enclave execution units.
In the optimistic case, when all parties behave honestly, \sysname requires only on-chain transactions for the creation of a \sysname contract as well as the locking and unlocking of user funds.
The smart contract execution itself is done without any on-chain transactions.

\subsection{Architecture Overview}
\begin{figure}[t]
	\centering
	\includegraphics[width=1\columnwidth, trim=0.7cm 0.08cm 0.6cm 0.1cm, clip]{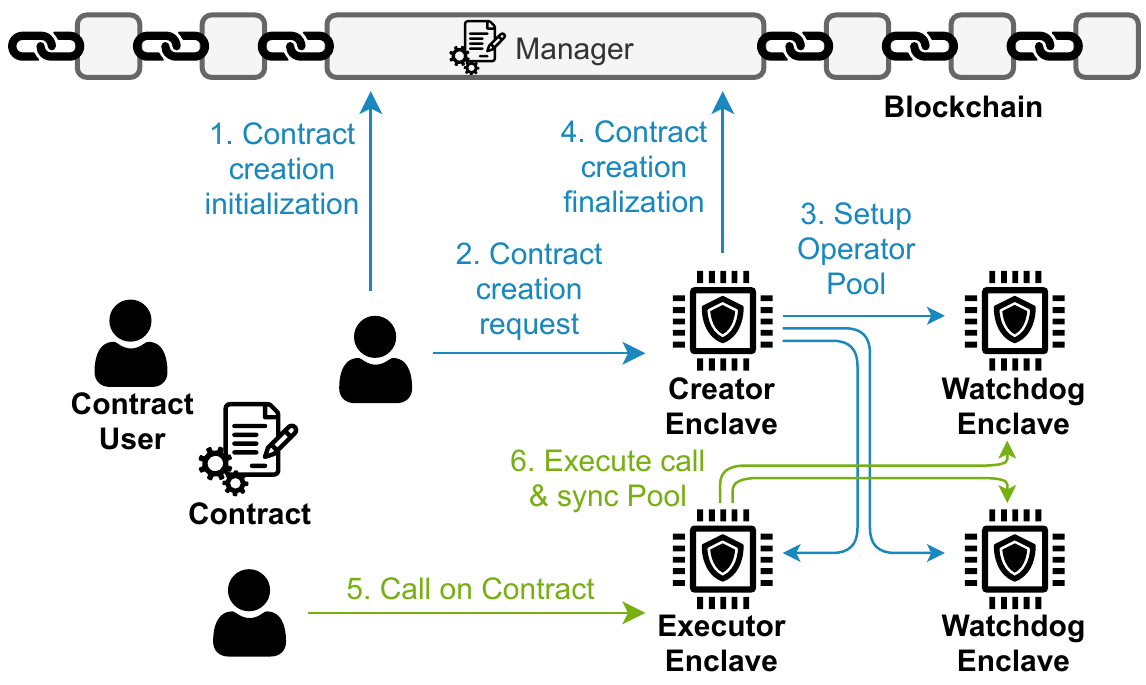}
	\caption{Exemplary overview how \sysname contracts are created (in blue) and executed (in green).}
	\label{fig:overview}
\end{figure}

\Cref{fig:overview} illustrates the high-level working of \sysname.
Before contract creation, there is already a set of enclaves that are registered with the on-chain manager contract.
The registration process is explained in detail in Section~\ref{sec:high level protocol/registration}.
To create a \sysname contract, a user will initialize a contract creation with the manager (Step 1), which includes a chosen enclave---out of the registered set---to execute the off-chain contract creation.
In Step 2, the chosen \emph{creator} enclave will setup the \emph{execution pool} for the given smart contract.
In \Cref{fig:overview}, the pool size is set to three; thus, the \emph{creator} enclave will randomly select three enclaves from the set of all enclaves registered in the system (Step 3).
In Step 4, the \emph{creator} enclave will submit the finalized contract information to the manager.
This includes the composition of the execution pool, i.e., a selected \emph{executor} enclave, which is responsible for executing the \sysname contract, as well as the \emph{watchdogs}, ensuring availability.
We elaborate on this in-depth in Section~\ref{sec:high level protocol/creation}.
In Step 5, another user can now call the new contract by directly contacting the executor.
Finally, for Step 6, the executor will execute the user's contract call and distribute the resulting state to the watchdog enclaves, which confirm the state update.
See Section~\ref{sec:high level protocol/execution} for a detailed specification of the execution protocol. 
If one of the enclaves stops participating (e.g., due to a crash), the dependent parties can challenge the enclave on the blockchain (see Section~\ref{sec:high level protocol/challenge-response}).
The dependent party can either be the user awaiting response from the executor or the executor waiting for the watchdogs' confirmation.
For example, if the executor stops executing the contract, the executor is challenged by the user.
A timely response constitutes a successful state transition as requested by the user.
Otherwise, if the current executor does not respond, one of the watchdogs will fill in as the new executor.
This makes \sysname highly available, as long as at least one watchdog enclave is dependable; thus, avoiding the need for collateral to incentivize correct behavior.
Further, \sysname supports private state, as the state is only securely shared with other enclaves.

\subsection{Design Challenges}
\label{sec:design-challenges}

We encountered a number of challenges while designing \sysname.
We briefly discuss them below.

\noindent\textbf{Protection Against Malicious Operators.}
\sysname's creator, executor, and watchdogs are protected in isolated enclaves running within the system, which is itself still under control of a potentially malicious operator.
Hence, operators can provide arbitrary inputs, modify honest users' messages, execute replay attacks, and withhold incoming messages.
Moreover, the system and its TEE (i.e., enclaves) can be turned off completely by its operator.
In order to protect honest users from malicious operators, we incorporate several security mechanisms.
While malicious inputs and modification of honest users' messages can easily be prevented using standard measures like a secure signature scheme, preventing withholding of messages is more challenging.
One particular reason is that for unreceived messages, an enclave cannot differentiate between unsent and stalled messages by the operator.
Hence, we incorporate an on-chain challenge-response procedure, which provides evidence about the execution request and the existence of a response to the enclave.

\noindent\textbf{Achieving Strong Liveness Guarantees.}
We enable dependent parties to challenge unresponsive operators via the blockchain.
The challenged operators either provide valid responses over the blockchain that dependent parties can use to finalize the state transition, or they are dropped from the execution pool.
In case an executor operator has been dropped, we use the execution pool to resume the execution; this requires state updates to be distributed to all watchdogs.
With at least one honest operator in the execution pool, the pool will produce a valid state transition.
Our protocol tolerates a fixed fraction of malicious operators as stated in our adversary model (cf. Section~\ref{sec:adversary-model}).
By selecting the pool members randomly, we guarantee with high probability that at least one enclave---controlled by an honest operator---is part of the execution pool.
We show in Section~\ref{sec:protocol-security} that our protocol achieves strong liveness guarantees.

\noindent\textbf{Synchronization with the Blockchain.}
Some of the actions taken by an enclave depend on blockchain data, e.g., deposits made by clients.
Hence, it is crucial to ensure that the blockchain data available to an enclave is consistent and synchronized with the main chain.
As an enclave does not necessarily have direct access to the (blockchain) network, it has to rely on the blockchain data provided by the operator.
However, the operator can tamper with the blockchain data and, e.g., withhold blocks for a certain time.
Thus, a major challenge is designing a synchronization mechanism that (i) imposes an upper bound on the time an enclave may lag behind the main chain, (ii) prevents an operator from isolating an enclave onto a fake side-chain, and (iii) ensures correctness and completeness of the blockchain data provided to the enclave, without (iv) requiring the enclave to validate or store the entire blockchain.
We present our synchronization mechanism addressing these challenges in Section~\ref{sec:high level protocol/synchronization}.

\noindent\textbf{Reducing Blockchain Interactions.}
Our system aims to minimize the necessary blockchain interactions to avoid expensive on-chain computations.
In the optimistic scenario, the only on-chain transactions necessary are the contract creation and the transfer of coins.
The transfer transactions can also be bundled to further reduce blockchain interactions.
Note that the virtualization paradigm known from state channels~\cite{PerunStateChannels} can be applied to our system.
This enables parties to install virtual smart contracts within existing smart contracts, and hence, without any on-chain interactions at all.
In the pessimistic scenario, i.e., if operators fail to provide valid responses, they have to be challenged, which requires additional blockchain interactions.

\noindent\textbf{Support of Private State.}
To support private state of randomized contracts, careful design is required to avoid leakage.
While the confidentiality guarantees of TEEs prevent any data leakage during contract execution, our protocol needs to ensure that an adversary cannot learn any information except the output of a successful execution.
In particular, in a system where the contract state is distributed between several parties, we need to prevent the adversary from performing an execution on one enclave, learning the result, and exploiting this knowledge when rolling back to an old state with another enclave.
This is due to the fact that a re-execution may use different randomness or different inputs resulting in a different output.
We prevent these attacks by outputting state updates to the users only if all pool members are aware of the new state.
Moreover, by solving the challenge of synchronization between enclaves and the blockchain, we prevent an adversary from providing a fake chain to the enclave, in which honest operators are kicked from the execution pool.
Such a fake chain would allow an attacker to perform a parallel execution.
While results of the parallel (fake) execution cannot affect the real execution, they can prematurely leak private data, e.g. the winner in a private auction.

\section{Definitions \& Notations}
\label{sec:definitions}

In the following, we introduce the cryptography primitives, definition, and notations used in the \sysname protocol.

\textbf{Cryptographic Primitives.}
\label{sec:protocol/crypto primitives}
Our protocol utilizes a public key encryption scheme $(\GenPK, \Enc, \Dec)$, a signature scheme $(\GenSig, \Sign, \Verify)$, and a secure hash function $\Hash(\cdot)$.
All messages sent within our protocol are signed by the sending party.
We denote a message $m$ signed by party $P$ as $(m;P)$.
The verification algorithm $\Verify(m')$ takes as input a signed message $m' \define (m;P)$ and outputs $\msgOk$ if the signature of $P$ on $m$ is valid and $\msgBad$ otherwise.
We identify parties by their public keys and abuse notation by using $P$ and $P$'s public key $\pk_P$ interchangeably.
This can be seen as a direct mapping from the identity of a party to the corresponding public key.

\textbf{TEE.} \label{sec:protocol/TEE} We comprise the hardware and software components required to create confidential and integrity-protected execution environments under the term TEE.
An operator can instruct her TEE to create new \emph{enclaves}, i.e., new execution environments running a specified program.
We follow the approach of Pass et al.~\cite{PassST17} to model the TEE functionality.
We briefly describe the operations provided by the ideal functionality formally specified in \cite[Fig. 1]{PassST17}.
A TEE provides a $\TEE.\install(\prog)$ operation which creates a new enclave running the program $\prog$.
The operation returns an enclave id $\txt{eid}$.
An enclave with id $\txt{eid}$ can be executed multiple times using the $\TEE.\resume(\txt{eid}, \txt{inp})$ operation.
It executes $\prog$ of $\txt{eid}$ on input $\txt{inp}$ and updates the internal state.
This means in particular that the state is stored across invocations.
The $\resume$ operation returns the output $\txt{out}$ of the program.
We slightly deviate from Pass et al.~\cite{PassST17} and include an attestation mechanism provided by a TEE that generates an attestation quote $\rho$ over $(\txt{eid}, \prog)$.
$\rho$ can be verified by using method $\VerifyQuote(\rho)$.
We consider only one instance $\enclave$ running the \sysname program per TEE.
Therefore, we simplify the notation and write $\enclave(\txt{inp})$ for $\TEE.\resume(\txt{eid}, \txt{inp})$.

\textbf{Blockchain.} \label{sec:protocol/blockchain}
We denote the blockchain by $\ledger$ and the average block time by $\blockTime$.
A block is considered final if it has at least $\nbConfBlocks$ confirmation blocks.
Throughout the protocol description in Section~\ref{sec:protocol/description}, enclaves consider only transactions included in final blocks.
Finally, we define that any smart contact deployed to the blockchain is able to access the current timestamp using the method $\ledger.\latestBock$ and the hash of the most recent 265  blocks~\cite{solidity} using the method $\ledger.\blockHash(i)$ where $i$ is the number of the accessed block. These features are available on Ethereum.

\section{The \sysname Protocol}
\label{sec:protocol}

The \sysname protocol considers four different roles: a manager smart contract deployed to the blockchain, operators that run TEEs, enclaves that are installed within TEEs, and users that create and interact with \sysname contracts.
In the following, we will shortly elaborate on the on-chain smart contract and the program executed by the enclaves, explain the \sysname protocol, and finally explain further security mechanisms that are omitted in the protocol description.

\subsection{Manager}
We utilize an on-chain smart contract in order to manage the \sysname system's on-chain interactions.
We call this smart contract \emph{manager} and denote it by $\manager$.
On the one hand, $\manager$ keeps track of all registered \sysname enclaves.
This enables the setup of an execution pool whenever an off-chain smart contract instance is created.
On the other hand, it serves as a registry of all \sysname contract instances.
$\manager$ stores parameters about each contract to determine the instance's status.
We denote the tuple describing a contract with identifier $\conId$ as $\manager^\conId$.
In particular, the manager stores the creator enclave ($\conCreator$), a hash of the program code ($\conCodeHash$), the set of enclaves forming the execution pool ($\conPool$), a total amount of locked coins ($\conBalance$), and a counter of withdrawals ($\conPayoutLevel$).
We set the field $\conCreator$ to $\bot$ after the creation process has been completed to identify that a contract is ready to be executed.
Moreover, for both executor and watchdog challenges, the contract allocates storage for a tuple containing the challenge message ($\conExecChalMsg$ resp. $\conWatchChalMsg$), responses ($\conExecChalRes$ resp. $\conWatchChalRes$), and the timestamp of the challenge submission ($\conExecChalBlock$ resp. $\conWatchChalBlock$).
A non-empty field $\conExecChalBlock$ resp. $\conWatchChalBlock$ signals that there is a running challenge.

Every \sysname enclave maintains a local version of the manager state extracted from the blockchain data it receives from the operator when being executed.
This enables all enclaves to be aware of on-chain events, e.g., ongoing challenges.

\subsection{\sysname Program}
\label{sec:protocol/enclave program}

All enclaves registered within the system run the \sysname program that enforces correct execution and creation of \sysname contracts.
In practice, the \sysname program's source code will be publicly available, e.g., in a public repository, so that the community can audit it.
Our protocol ensures that all registered enclaves run this code using remote attestation (cf. Section~\ref{sec:high level protocol/registration}: Enclave registration).
We present methods required for the execution protocol in Program~\ref{prg:program execution} and defer methods for the contract creation into Program~\mbox{\ref{prg:program specification 2}} in Appendix \mbox{\ref{sec:app-further-protocol}}.

Whenever an enclave is invoked, it synchronizes itself with the blockchain network and receives the relevant blockchain data in a reliable way (cf. Section~\ref{sec:high level protocol/synchronization}).
This way, the POSE program has access to the current state of the manager.
In order to support arbitrary contracts, we define a common interface in Section~\ref{sec:protocol/smart contracts} that is used by the POSE program to invoke contracts.

Enclaves running the \sysname program only accept signed messages as input.
The public keys of pool members for signature verification are derived from the synchronized blockchain data.
According to our adversary model (cf. Section~\ref{sec:adversary-model}), the adversary cannot read or tamper messages originating from honest users or the enclave itself.
Further, the contracts themselves keep track of already received execution  requests and do not perform state transitions for duplicated requests. (cf. Section~\ref{sec:protocol/smart contracts}).
This prevents replay attacks against both, executive and watchdog enclaves.

\begin{boxProgram}[float, label={prg:program execution}]{\sysname Program (execution) executed by enclave $T$}
	\raggedright
	Upon invocation with input blockchain data $\ledger$, store $\ledger$.
	
	Upon receiving $m \define (\msgExecute,\conId,r,\move;\user)$, do:
	\begin{enumerate}
		\item If $\manager^{\conId}.\conPool[0] \neq \tee$ or $\openResponses^{\conId} \neq \emptyset$, return $(\msgBad)$.
		
		\item Execute $C_{\conId}.\conNextState(\user,\ledger,\move, \Hash(m))$.
		
		\item Store $\openResponses^{\conId} = \manager^{\conId}.\conPool$ and $h^{\conId} = \Hash(m)$, set $c = \Enc(C_{\conId}.\conGetState(\txt{all});\key^{\conId})$ and return $(\msgUpdate,\conId, c, h^{\conId};\tee)$.
		
	\end{enumerate}
	
	Upon receiving $m \define (\msgUpdate, \conId, c, h;\tee')$, do:
	\begin{enumerate}
		\item If $\tee' \neq \manager^{\conId}.\conPool[0]$ or $\tee \notin \manager^{\conId}.\conPool$, return $(\msgBad)$.
		
		\item Define $\state = \Dec(c;\key^{\conId})$ and call $C_{\conId}.\conUpdateState(\state, h)$.
		
		\item Return $(\msgConfirm,\conId,h;\tee)$.
	\end{enumerate}
	
	Upon receiving $\{m_i \define (\msgConfirm,\conId,h_i; \tee_i)\}_i$, do:
	\begin{enumerate}
		\item If $\manager^{\conId}.\conPool[0] \neq \tee$ or $\openResponses^{\conId} = \emptyset$, return $(\msgBad)$.
		
		\item Set $\openResponses^{\conId} = \openResponses^{\conId} \cap \manager^{\conId}.\conPool$.
		
		\item For each $m_i$ do:
		\begin{itemize}
			\item If $h_i \neq h^{\conId}$ or $\tee_i \notin \openResponses^{\conId}$, skip $m_i$.
			\item Otherwise remove $\tee_i$ from $\openResponses^{\conId}$.
		\end{itemize}
		
		\item If $\openResponses^{\conId} \neq \{\tee\}$, return $(\msgBad)$.
		Otherwise, set $\openResponses^{\conId} = \emptyset$, $\state \define C_{\conId}.\conGetState(\txt{pub})$ and return $(\msgOk, \conId, \state,h^{\conId};\tee)$.
	\end{enumerate}
	
\end{boxProgram}

\subsection{\sysname Contracts}
\label{sec:protocol/smart contracts}

Although our system supports the execution of arbitrary smart contracts, the contracts need to implement a specific interface
(cf. Program~\ref{prg:contract-interface}).
This allows any \sysname enclave to trigger the execution without knowing details about the smart contract functionality.
Upon an execution request from some user, the \sysname enclave provides the user's identity $\user$, blockchain data $\ledger$, the description of the user's request, $\move$, and the request hash, $h$, to the smart contract's method $\conNextState$.
The smart contract first processes the relevant blockchain data and marks the current length of the blockchain as processed.
This feature is mainly used to enable smart contracts to deal with money, i.e., to detect on-chain deposits and withdrawals.
We elaborate on the processing of blockchain data in Section~\ref{sec:high level protocol/synchronization}, and on the money mechanism of the \sysname system in Appendix~\ref{sec:protocol/payments}.

	Note that double spending within a contract is prevented due to sequential processing of any execution request, and double spending of on-chain payouts is prevented by the mechanism explained in Appendix~\ref{sec:protocol/payments}.
After the blockchain data is processed, $\conNextState$ executes the move requested by the user and updates the state accordingly.
Method $\conUpdateState$ takes state $\txt{new}$ and hash $h$ (for preventing replay attacks) as input and sets $\txt{new}$ as the contract state.
This includes the length of the blockchain that is marked as processed.
Further, the smart contract provides method $\conGetState$.
If called with $\flag = \txt{all}$, it returns the whole smart contract state.
Otherwise, if called  with $\flag = \txt{pub}$, it returns only the public state.
In order to prevent replay attacks, each smart contract maintains a list with the hashes of already received execution requests, $\received$.
In case of duplicated requests, i.e., $h \in \received$, both the $\conNextState$ method and the $\conUpdateState$ method, do not perform any state transition.
Instead, they interpret the request as a dummy move that has no effect on the state.
If executed successfully, the $\conNextState$ method adds the executed request to $\received$, i.e., $\received = \received \cup \{h\}$.
As $\received$ is part of the state, it is updated by the $\conUpdateState$ method as well.
While it might seem counter intuitive to overwrite the list of received requests, this feature is required to ensure that all enclaves are aware of the same transition history; even if an executor distributes a state update to just a subset of watchdogs before getting kicked~\footnote{In practice, the state update removes at most the last element from the request history; a fact that can be exploited to reduce the size of state updates.}.

We consider the initial state of a smart contract to be hard-coded into the smart contract description.
If an enclave creates a new smart contract instance, the initial state is automatically initialized.
A contract state additionally contains a variable to store the highest block number of the already processed blockchain data.
This variable is used to detect which transactions of received blockchain data have already been handled.

\begin{boxProgram}[float, label={prg:contract-interface}, fontupper=\small]{Interface of a contract $C$ executed within a \sysname enclave}
	Function: $\conNextState(\user, \ledger, move, h)$

	Function: $\conUpdateState(new, h)$
	
	Function: $\conGetState(flag)$
\end{boxProgram}

\subsection{Synchronization}
\label{sec:high level protocol/synchronization}

As some of the actions taken by an enclave depend on blockchain data, e.g., deposits to the contract, it is crucial to ensure that the blockchain state available to a registered enclave $\enclave$ is consistent and synchronized with the main chain.
In particular, blocks that are considered final by some party, will eventually be considered final by all parties.
We design a synchronization mechanism that allows $\enclave$ to synchronize itself without having to validate whole blocks.
Note that $\enclave$ has access to a relative time source according to our adversary model (see Section~\ref{sec:adversary-model}).

Upon initialization, $\enclave$ receives a chain of block headers $\ledgerHeaders$ of length $\nbConfBlocks + 1$.
Note that the first block $\checkpoint$ of $\ledgerHeaders$ can be considered final since it has $\nbConfBlocks$ confirmation blocks.
First, $\enclave$  checks that $\ledgerHeaders$ is consistent in itself and sets its own clock to be the one of the latest block's timestamp.
Second, $\enclave$ signs block $\checkpoint$ as blockchain evidence that needs to be provided to the manager.
The registration mechanism (cf. Section~\ref{sec:high level protocol/registration}) uses this evidence to ensure that $\enclave$ has been initialized with a valid sub-chain of the main-chain up to block $\checkpoint$.
Further, the registration mechanism checks that $\checkpoint$ is at most $\tau^{\txt{on}}_{\txt{slack}}$ blocks behind the current one;
$\tau^{\txt{on}}_{\txt{slack}}$  needs to account for the confirmation blocks and the fact that transactions are not always mined immediately.
Via this parameter, we can set an upper bound to the time $\tau^{\txt{off}}_{\txt{slack}}$ an enclave may lag behind; $\tau^{\txt{off}}_{\txt{slack}}$ additionally considers potential block variance and the fact that miners have some margin to set timestamps.
In the following, we call $\tau^{\txt{off}}_{\txt{slack}}$  \textit{slack}~\footnote{We can reduce the slack assuming an absolute source of time realized via trusted NTP servers, cf.~\cite{ekiden}, by enabling the enclave to check if she was invoked with the most recent block headers up to some variance of the timestamps.}.
Clients that want a contract execution to capture on-chain effects, e.g., deposits, wait until the enclave considers the corresponding block as final, even when being at slack.

Once successfully initialized, $\enclave$ synchronizes itself with the blockchain.
Whenever a registered enclave is executed throughout the protocol, it receives the sub-chain of block headers $\ledgerHeaders'$ that have been mined since the last execution.
$\enclave$ checks that $\ledgerHeaders'$ is a valid successor of $\ledgerHeaders$ where blocks in $\ledgerHeaders$ that have not been final may change.
Further, $\enclave$ checks that the latest block in $\ledgerHeaders'$ is at most $\tau_{variance}$ behind the own clock; $\tau_{variance}$  captures the variance in the block creation time and the fact that miners have some margin to set timestamps.
When receiving a block that is before the own clock, the clock is adjusted.

Finally, we need to prevent an operator from isolating its enclave by setting up a valid sidechain with manipulated timestamps.
To this end, we require the operators to periodically provide new blocks to $\enclave$ even if $\enclave$ does not need to take any action.
In particular, we require that the operator provides at least $L$ blocks within time $\tau_{p}$ where $\tau_p$ accounts for potential block time variances.
The system is secure as long as the attacker cannot mine $L$ blocks within time $\tau_{p}$ while the honest miners can.
Hence, the selection of $\tau_{p}$ and $L$ has some implications on the fraction of adversarial computing power that can be tolerated by the system.
Since 2018, an interval of 50 (100, 200, 300) blocks took at most 33 (28, 26, 25) seconds per block \cite{google-cloud-big-query-block-variance}, which might all be reasonable choices for $L$ and $\frac{\tau_{p}}{L}$.
As the average block time is around 13 seconds \cite{etherscan-blocktime}, the adversary gets $2-3$ times more time to mine the blocks of its sidechain.
This means that the system can tolerate adversarial fractions from a third (when instantiated with $L = 300$ and $\tau_{p} = 25 \cdot L$) to a forth (when instantiated with $50$ and $33 \cdot L$).

While the above techniques allow an enclave to synchronize itself, the enclave does not have access to the block data, yet.
Instead of requiring enclaves to validate whole blocks, we require operators to filter the relevant transactions and provide them to the enclave while enabling the enclaves to check correctness and completeness of the received data itself.
For the latter, we introduce $\txt{incrTxHash}$, a hash maintained by the manager and all initialized enclaves that is based on all relevant transactions.
Whenever the manager receives a relevant transaction $tx$, it updates $\txt{incrTxHash}$, such that  $\txt{incrTxHash}_{i+1}$ is defined as
\begin{equation*}
H(\txt{incrTxHash}_i~||~tx.data~||~tx.sender~||~tx.value)
\end{equation*}
where $tx.data$ is the raw data of $tx$, $tx.sender$ denotes the creator of $tx$, and $tx.value$ contains the amount of any deposits or withdrawals.
Whenever enclaves are invoked with new blocks, operators additionally provide all relevant transactions.
This way, enclaves can re-compute the new incremental hash and compare the result to the on-chain value of $\txt{incrTxHash}$.
In order to verify that the on-chain $\txt{incrTxHash}$ is indeed part of the main chain, operators additionally provide a Merkle proof showing that $\txt{incrTxHash}$ is part of the state tree.
The proof can be validated using the state root, which is part of the block headers provided to the enclaves.
This way, enclaves can ensure that operators have not omitted or manipulated any relevant transactions.

\subsection{Protocol Description}
\label{sec:protocol/description}

In this section, we dive into a detailed description of our protocol.
We present 1) enclave registration, 2) contract creation, 3) contract execution, and 4) the challenge-response parts of our protocol.
The \sysname program running inside the operators' enclaves is stated in Section~\ref{sec:protocol/enclave program}.
For the sake of exposition, we extracted the validation steps performed by the manager on incoming messages into Program~\ref{prg:verify-program} in Appendix~\ref{sec:app-further-protocol}.
Further, we elaborate in Appendix~\ref{sec:protocol/payments} on the coin flow within the protocol.

\subsubsection{Enclave Registration}
\label{sec:high level protocol/registration}

Operator $O$ controlling some TEE unit can contribute to the \sysname system by instructing his TEE to create a new \sysname enclave $\enclave_O$.
The protected execution environment $\enclave_O$ needs to be initialized with the \sysname program presented in Section~\ref{sec:protocol/enclave program}.
During the creation of $\enclave_O$, an asymmetric key pair $(\pk_O, \sk_O)$ is generated.
The secret key $\sk_O$ is stored inside the enclave and hence is only accessible by the \sysname program running in $\enclave_O$.
The public key $\pk_O$ is returned as output to the operator.
Furthermore, operator $O$ uses the TEE to produce an attestation $\rho_O$ stating that the freshly generated enclave $\enclave_O$ runs the \sysname program and controls the secret key corresponding to $\pk_O$.\footnote{An attestation mechanism can be designed based on a chain of trust, where the TEEs manufacturer's public key represents the root.
	This way a smart contract knowing a list of public keys can verify an attestation quote without further interaction.
	We omit further details about the practical implementation and refer the reader to~\cite{PassST17}.}

Finally, $O$ sends the latest $\nbConfBlocks + 1$ block headers $\ledgerHeaders$ together with the relevant blockchain data to the enclave which validates the consistency of the block headers and completeness of the blockchain data (cf. Section~\ref{sec:high level protocol/synchronization}) and returns a blockchain evidence $\rho^\ledger_O$, i.e.,
a signed tuple containing the blockhash and the number of the latest final block known to the enclave.
After operator $O$ created a new \sysname enclave $\enclave_O$, $O$ can register $\enclave_O$ by sending $m \define$ $(\msgRegister, \enclave_O, \rho_{O}, \rho^\ledger_O; O)$ to manager $\manager$.
$\manager$ verifies that $\rho_{O}$ is a valid attestation and that $\rho^\ledger_O$ refers to a block on the blockchain known to $\manager$ that is not older than $\tau^{\txt{on}}_{\txt{slack}}$ blocks.
If the check holds and the signature of the operator is valid, i.e., $\Verify(m) = \msgOk$,
$\manager$ adds $\enclave_O$ (identified by its public key $\pk_O$) to the set of registered enclaves, i.e., $\manager.\tees \define \manager.\tees \cup \{\enclave_O\}$.
This procedure ensures that all registered enclaves run the \sysname program and that the secret key $\sk_O$ remains private.
Hence, re-attesting enclaves during later protocol steps is not needed.

\subsubsection{Contract Creation}
\label{sec:high level protocol/creation}

The creation protocol is initiated by a user $\user$ who wants to install a new smart contract, with program code $\code$, into the \sysname system.
We outline the protocol in the following and provide a full explanation and specification in Appendix~\mbox{\ref{sec:app-further-protocol}}.

$\user$ picks an arbitrary registered enclave $\enclave_C$ and sends a creation initialization to $\manager$ containing $\Hash(\code)$ and $\enclave_C$.
The manager $\manager$ allocates a new contract tuple with a fresh identifier $\conId$.
Next, $\user$ sends a creation request, containing $\code$, to $\enclave_C$ which randomly selects $\poolsize$ enclaves for the contract execution pool and samples a symmetric pool key.
The generated information is distributed in a confidential way to all pool enclaves, which install a new smart contract with code $\code$ and confirm the installation to $\enclave_C$.
Finally, $\enclave_C$ signs a creation confirmation, which is submitted to $\manager$ that marks the contract as created.

If the contract is not created within a certain time, $\user$ starts a creation challenge.
If any pool member does not respond to $\enclave_C$ timely, $\enclave_C$ starts a pool challenge (cf. \Cref{sec:high level protocol/challenge-response}).

\subsubsection{Contract Execution}
\label{sec:high level protocol/execution}

\begin{figure}
	\begin{center}
		\begin{tikzpicture}
		
		\newcommand{\rightMost}{8.25};
		\newcommand{\leftMost}{-0.25};
		
		\tikzset{header/.style={draw, font=\bf, fill = white, below, align=center, minimum width=2.5cm,minimum height=0.7cm}}
		\tikzset{event/.style={draw, font=\tiny, fill=white, dashed, below, align=center, minimum width=2cm,minimum height=0.5cm}}
		\tikzset{arrowLabel/.style={draw, draw = white, align=center, fill=white,font=\tiny}}
		\tikzset{event-right/.style={draw, font=\tiny, fill=white, dashed, below left, align=right, minimum width=2cm,minimum height=0.5cm}}
		\tikzset{event-left/.style={draw, font=\tiny, fill=white, dashed, below right, align=left, minimum width=2cm,minimum height=0.5cm}}
		\tikzset{event-enter/.style={draw, circle, line width=0.25mm, fill=white,inner sep=1.5pt}}

		\draw (1,-7.25) -- (1, 0.5);
		\draw (4,-7.25) -- (4, 0.5);
		\draw (7,-7.25) -- (7, 0.5);
		
		\node[header] (U) at (1,0.8) {User: $\user$};
		\node[header] (EO) at (4,0.8) {Executer: $E$ \\
									\tiny {($\define \manager^{\conId}.\conPool[0]$)}};
		\node[header] (WO) at (7,0.8) {Watchdog: $W$ \\
									\tiny {($\in \manager^{\conId}.\conPool \setminus E$)}};
		
		\node[event-left] at (\leftMost,-0.2) {On input $(\conId,\move)$\\
			$r \in_R \{0,1\}^\kappa$};
		
		\node[event-enter] at (1, -0.2){};
		
		\draw[->] (1,-1.2) -- node[arrowLabel] {$m = (\msgExecute,\conId,$\\$r,\move; \user)$} (4,-1.2);
		
		\node[event] at (4,-1.5) {Finalize current and pending\\
			executions or challenges corresponding to $\conId$.\\\\
			$\pre := \tee_E(\ledger,m)$.
		};
		\node[event-enter] at (4, -1.5){};
		
		\draw[->] (4,-2.9) -- (7,-2.8);
		\draw[->] (4,-2.9) -- (7,-3.0);
		\draw[->] (4,-2.9) -- node[arrowLabel] {$(\pre)$} (7,-2.9);
		
		\node[event-right] at (\rightMost,-3.3) {$\conf_W \define \tee_W(\ledger, \pre)$\\
			If $\txt{conf}_W = (\msgBad, \cdot)$, abort.};
		\node[event-enter] at (7, -3.3){};
		
		\draw[->] (7,-4.2) -- (4,-4.2);
		\draw[->] (7,-4.4) -- (4,-4.4);
		\draw[->] (7,-4.3) -- node[arrowLabel] {$(\conf_W)$} (4,-4.3);
		
		\node[event] at (4,-4.7) {After time $\timeOffchainPropagation,$ 
			$\res = \tee_E(\ledger, \{\conf_W\})$.\\
			If $\res = (\msgBad)$, 
			$\res = \protWatchdogChallenge(\pre)$.
		};
		\node[event-enter] at (4,-4.7){};
		
		\draw[->] (4,-5.7) -- node[arrowLabel] {$\res = (\msgOk, \conId, \state, h; \tee_E)$} (1,-5.7);
		
		\node[event-left] at (\leftMost,-6.0) {In time $\timeOffchainExecution$ after sending $m$:\\
			If $\res$ has not been received, $\Verify(\res) = \msgBad$ or  $h \ne \Hash(m)$,\\
			execute $\res \define \protExecutiveChallenge(m)$.\\
			If $\res = (\msgBad)$ and $\manager^{\conId}.\conPool \ne \emptyset$, restart execution with same $r$.};
		\node[event-enter] at (1,-6.0){};
		
		\end{tikzpicture}
	\end{center}
	\caption{Detailed execution protocol.}
	\label{fig:diagram-execution}
\end{figure}
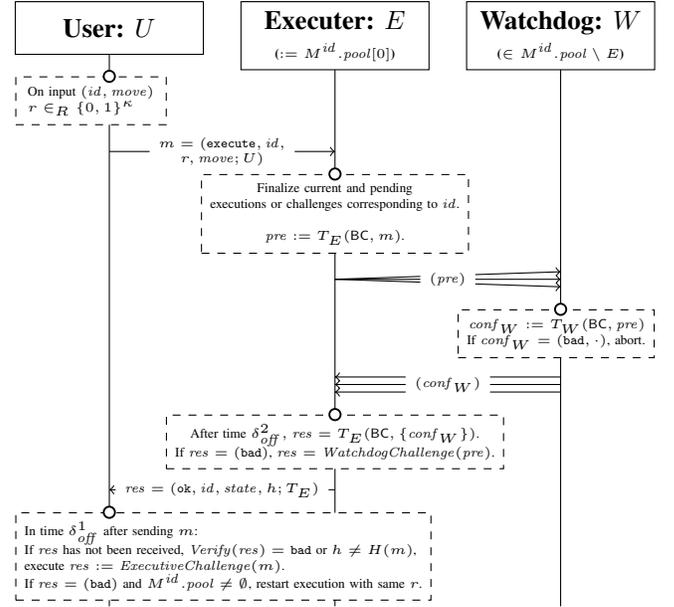

The execution protocol is initiated by a user $\user$ who wants to execute an existing smart contract, identified by $\conId$, with input $\move$.
The protocol is specified in Figure~\ref{fig:diagram-execution}.
Program \ref{prg:program execution} specifies the parts of the \sysname program that are relevant for the contract execution.

To trigger the execution, $\user$ sends an execution request to operator $E$ controlling the executor enclave $\enclave_E$, the first enclave in the contract pool stored at $\manager$.
$\enclave_E$ executes the request and securely propagates the new state to all other pool members, called watchdogs.
If any watchdog does not confirm in time, it is challenged by $E$ (cf. \textit{Challenge-Response}).
Eventually, $\enclave_E$ receives confirmations from all watchdogs or the unresponsive watchdogs are kicked out of the pool.
Either way,  $\enclave_E$ outputs the new public state to $\user$.
We want to stress that this way no party gets to know the result of an update before all pool members agree on the update.
If $E$ does not respond in time, it is challenged by $\user$ (cf. \textit{Challenge-Response}).
If $E$ does not respond to the challenge, it is kicked from the pool by $\user$.
The next enclave in the pool, $\enclave_E'$, takes over as the new executor.
At this point, the new executor might be on a different state than the other pool members, since $\enclave_E'$ might have received the previous state update but some other pool members not, or vice versa.

Our system automatically ensures that all enclaves share the same contract state after the next successful execution, in which  $\enclave_E'$ distributes its state to the other enclaves.
Let us call the previous incompletely distributed update $\txt{update}$ and the new updated initiated by $\enclave_E'$ $\txt{update}'$.
In case $\enclave_E'$ has received $\txt{update}$, $\txt{update}'$ is a successor of $\txt{update}$, and hence, covers both updates.
This way, a watchdog that updates to $\txt{update}'$ essentially contains both executions, $\txt{update}$ and $\txt{update}'$.
In case $\enclave_E'$ has not received $\txt{update}$ but the other watchdogs have, $\enclave_E'$ either propagates the update already known to the watchdogs, i.e., $\txt{update} = \txt{update}'$, or a concurrent one, i.e., $\txt{update} \ne \txt{update}'$.
For the former, the watchdogs interpret the update as a dummy update without any effect as the corresponding execution request is already within their list of received request hashes (cf. Section~\ref{sec:protocol/smart contracts}).
For the latter, the update of the watchdogs is overwritten by the one of the executive enclave.
As $\txt{update}$ has been incomplete, and hence, produced no public output, it is safe to overwrite this update. 
To produce a public output for $\txt{update}$, all pool enclaves including $\enclave_E'$ would have to confirm $\txt{update}$.

Finally, $\user$ can just submit the previous execution request with the same random nonce $r$ to $\enclave_E'$.
In case the enclave has already seen this request, it is interpreted as empty dummy move which prevents a duplicated execution.

\subsubsection{Challenge-Response}
\label{sec:high level protocol/challenge-response}

\begin{figure}
	\begin{center}
		\begin{tikzpicture}

\newcommand{\rightMost}{8.25};
\newcommand{\leftMost}{-0.25};

\tikzset{header/.style={draw, font=\bf, fill = white, below, align=center, minimum width=2.5cm,minimum height=0.7cm}}
		\tikzset{event/.style={draw, font=\tiny, fill=white, dashed, below, align=center, minimum width=2cm,minimum height=0.5cm}}
\tikzset{arrowLabel/.style={draw, draw = white, align=center, fill=white,font=\tiny}}
\tikzset{event-right/.style={draw, font=\tiny, fill=white, dashed, below left, align=right, minimum width=2cm,minimum height=0.5cm}}
\tikzset{event-left/.style={draw, font=\tiny, fill=white, dashed, below right, align=left, minimum width=2cm,minimum height=0.5cm}}
\tikzset{event-enter/.style={draw, circle, line width=0.25mm, fill=white,inner sep=1.5pt}}

\draw (1,-7.8) -- (1,0);
\draw (4,-7.8) -- (4,0);
\draw (7,-7.8) -- (7,0);

\node[header] at (1,0.6) {User: $\user$};
\node[header] at (4,0.6) {Manager: $\manager$};
\node[header] (EO) at (7,0.6) {Executer: $E$ \\
	\tiny {($\define \manager^{\conId}.\conPool[0]$)}};

\draw[->] (0,-0.4) -- node[arrowLabel] {$m$} (1,-0.4);
\draw[->] (1,-0.6) -- node[arrowLabel] {$m = (\msgExecute,\conId,$\\$n,\move; \user)$} (4,-0.6);

\node[event] at (4,-0.9) {If $\Validate(1, m; \manager^{\conId}) = \msgBad$, discard.\\
	Set $\manager^{\conId}.\conExecChalMsg = m$, $\manager^{\conId}.\conExecChalBlock = \ledger.\latestBock$ and
	$\manager^{\conId}.\conExecChalRes = \bot$. 
};
\node[event-enter] at (4, -0.9){};

\draw[->] (4,-1.9) -- node[arrowLabel] {$(m)$} (7,-1.9);

\node[event-right] at (\rightMost,-2.2) {Handle $m$ like a message directly received by $U$\\
	until receiving $\res = (\msgOk, \dots)$ from $\tee_E$,\\
	but priortize it above other pendinging executions.
};
\node[event-enter] at (7, -2.2){};

\draw[->] (7,-3.5) -- node[arrowLabel] {$\res= (\msgOk, \conId,$\\$\state, h;\tee_E)$} (4,-3.5);

\node[event] at (4,-3.8) {
	If $\Validate(2, \res; \manager^{\conId}) = \msgBad$, discard.\\
	Set $\manager^{\conId}.\conExecChalMsg = \bot$, $\manager^{\conId}.\conExecChalBlock = \bot$ and
	$\manager^{\conId}.\conExecChalRes = \res$. 
};
\node[event-enter] at (4, -3.8){};

\draw[->] (4,-4.8) -- node[arrowLabel] {$(\res)$} (1,-4.8);

\draw[->] (1,-5) -- node[arrowLabel] {$(\res)$} (0,-5);

\node[event-left] at (\leftMost,-5.2) {If $(\res)$ has not been received\\
	within time $\timeOnchainExecution$ after sending $m$.};
\node[event-enter] at (1, -5.2){};

\draw[->] (1,-6.2) -- node[arrowLabel] {$(\msgFinalize, 1, \conId)$} (4,-6.2);

\node[event] at (4,-6.5) {
	If $\Validate(3;\manager^{\conId}) = \msgBad$, discard.\\
	Remove $\manager^{\conId}.\conPool[0]$ from $\manager^{\conId}.\conPool$
	and set $\manager^{\conId}.\conExecChalBlock \define \bot$.
};
\node[event-enter] at (4, -6.5){};

\draw[<-] (0,-7.5) -- node[arrowLabel] {$(\msgBad)$} (1,-7.5);

\end{tikzpicture}
	\end{center}
	\caption{Detailed executor challenge protocol.}
	\label{fig:diagram-challenge-executor}
\end{figure}
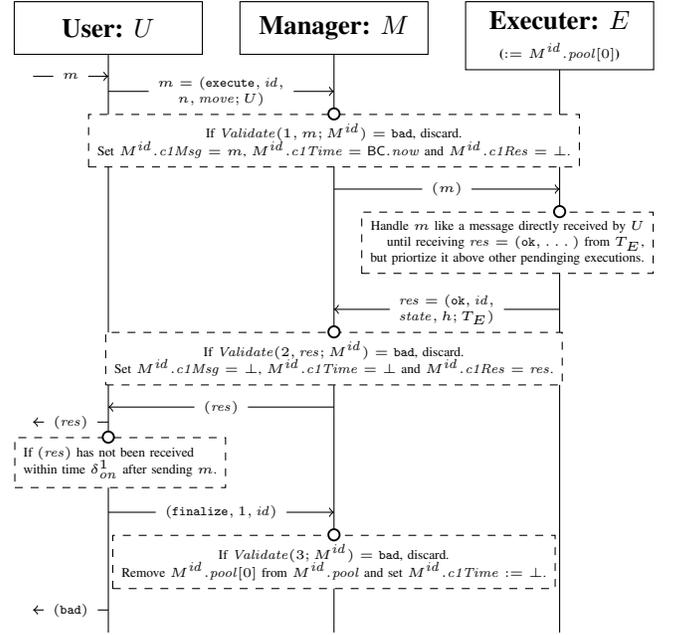

If any party does not receive a \textit{timely} response to its messages during the off-chain execution, it challenges the receiver on-chain.
Therefore, all operators need to monitor the blockchain for any on-chain challenges.
We will elaborate on the timeouts $(\delta^{\dag}_\star)$, where $\dag \in \{0,1\}$ and $\star \in \{\txt{off}, on\}$, which define the notion of \textit{timely} in Appendix~\ref{sec:high level protocol/timeouts}.
In particular, we describe the relation between $\delta^1_*$ and $\delta^2_*$.
The challenge-response procedure is executed in all of the following cases.
\begin{enumerate}[label=(\alph*)]
	\item The creator enclave has not responded to the user within time $\timeOffchainExecution$ during the contract creation protocol.
	\item At least one pool enclave has not responded to the creator enclave within time $\timeOffchainPropagation$ during the contract creation protocol.
	\item The executor enclave has not responded to the user within time $\timeOffchainExecution$ during the contract execution protocol.
	\item At least one watchdog enclave has not responded to the executor enclave within time $\timeOffchainPropagation$ during the contract execution protocol.
\end{enumerate}
Since (a) is conceptually identically to (c) and (b) to (d), we present the executor challenge (c) and the watchdog challenge (d) in Figure~\ref{fig:diagram-challenge-executor} and Figure~\ref{fig:diagram-challenge-watchdogs}.
The specifications of (a) and (b) are provided in the Appendix \mbox{\ref{sec:app-further-protocol}} in Figure~\mbox{\ref{fig:diagram-creation-challenge}} and Figure~\mbox{\ref{fig:diagram-creation-watchdog-challenge}}.

For the executor challenge as shown in Figure~\ref{fig:diagram-challenge-executor}, suppose user $\user$ has not received a result from the executor enclave $\enclave_E$ within time $\timeOffchainExecution$, then, $\user$ starts the challenge-response protocol.
To this end, $\user$ sends the execution request to the manager $\manager$ who verifies the validity of the message (cf. Program~\ref{prg:verify-program}).
If all checks hold, $\manager$ stores the challenge message and then starts timeout $\timeOnchainExecution$ by storing the current timestamp.
As soon as the challenge message is recorded on-chain, the operator of the executor enclave $\enclave_E$ extracts the execution request from the challenge and starts the execution.
Performing the execution request is identical to the standard execution as described in Section~\ref{sec:high level protocol/execution}.
However, the operator prioritizes challenges over off-chain execution requests to avoid getting kicked.
Additionally, if $\enclave_E$ already performed the state update and state propagation, the operator may use the already obtained result as response.
Either way, if the operator sends a response message in time, the manager $\manager$ checks the validity of the message and whether or not it matches the stored challenge.
If all checks succeed, $\manager$ stores the result and removes the challenge message.
This finalizes the challenge procedure.
If the operator does not send a valid response in time $\timeOnchainExecution$, user $\user$ sends message $\msgFinalize$ to $\manager$.
This triggers the manager to kick $\enclave_E$ from the execution pool of this contract and assign the next enclave in the list as the new executor enclave, if possible.
Then, if the pool is not empty, $\user$ restarts the execution.
As $\manager$ only accepts a response if the operator executed the challenged request correctly, the described procedure ensures that there is either a consistent state transition or $\enclave_E$ is kicked from the execution pool, hence, ensuring liveness as long as there remains one active operator.

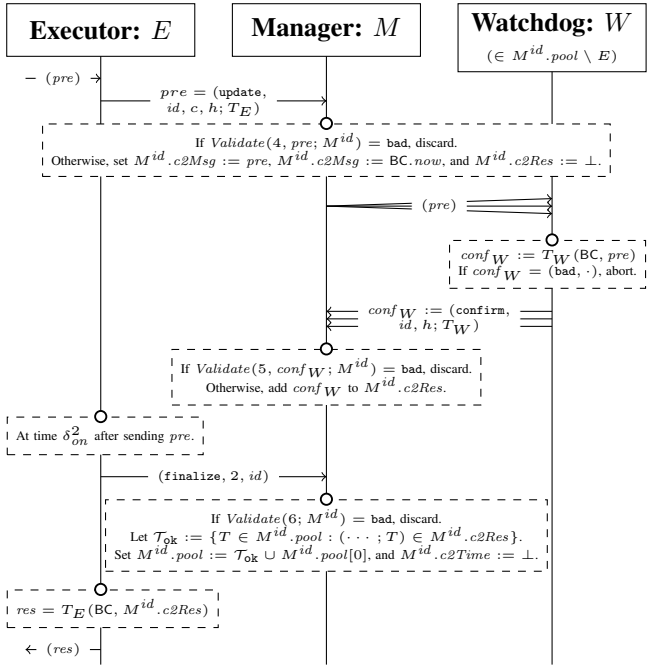
\begin{figure}
	\begin{center}
		\begin{tikzpicture}
\newcommand{\rightMost}{8.25};
\newcommand{\leftMost}{-0.25};

\tikzset{header/.style={draw, font=\bf, fill = white, below, align=center, minimum width=2.5cm,minimum height=0.7cm}}
		\tikzset{event/.style={draw, font=\tiny, fill=white, dashed, below, align=center, minimum width=2cm,minimum height=0.5cm}}
\tikzset{arrowLabel/.style={draw, draw = white, align=center, fill=white,font=\tiny}}
\tikzset{event-right/.style={draw, font=\tiny, fill=white, dashed, below left, align=right, minimum width=2cm,minimum height=0.5cm}}
\tikzset{event-left/.style={draw, font=\tiny, fill=white, dashed, below right, align=left, minimum width=2cm,minimum height=0.5cm}}
\tikzset{event-enter/.style={draw, circle, line width=0.25mm, fill=white,inner sep=1.5pt}}

\draw (1,-8.3) -- (1,0);
\draw (4,-8.3) -- (4,0);
\draw (7,-8.3) -- (7,0);

\node[header] at (1,0.5) {Executor: $E$};
\node[header] at (4,0.5) {Manager: $\manager$};
\node[header] at (7,0.5) {Watchdog: $W$\\
							\tiny{$(\in \manager^{\conId}.\conPool \setminus E)$}};

\draw[->] (0,-0.5) -- node[arrowLabel] {$(\pre)$} (1,-0.5);
\draw[->] (1,-0.8) -- node[arrowLabel] {$pre = (\msgUpdate,$\\$\conId,c,h;\tee_E)$} (4,-0.8);

\node[event] at (4,-1.1) {If $\Validate(4, \pre; \manager^{\conId}) = \msgBad$, discard.\\
	Otherwise, set $\manager^{\conId}.\conWatchChalMsg \define \pre$, $\manager^{\conId}.\conWatchChalMsg \define \ledger.\latestBock$, and $\manager^{\conId}.\conWatchChalRes \define \bot$.
};
		\node[event-enter] at (4, -1.1){};

\draw[->] (4,-2.2) -- (7,-2.1);
\draw[->] (4,-2.2) -- (7,-2.3);
\draw[->] (4,-2.2) -- node[arrowLabel] {$(\pre)$} (7,-2.2);

\node[event-right] at (\rightMost,-2.65) {$\conf_W \define \tee_W(\ledger, \pre)$\\
	If $\txt{conf}_W = (\msgBad, \cdot)$, abort.};
		\node[event-enter] at (7, -2.65){};

\draw[->] (7,-3.6) --(4,-3.6);
\draw[->] (7,-3.8) --(4,-3.8);
\draw[->] (7,-3.7) -- node[arrowLabel] {$\conf_W \define (\msgConfirm,$\\$\conId, h; \tee_W)$} (4,-3.7);

\node[event] at (4,-4.1) {If $\Validate(5, \conf_W; \manager^{\conId}) = \msgBad$, discard.\\
	Otherwise, add $\conf_W$ to $\manager^{\conId}.\conWatchChalRes$.
};
		\node[event-enter] at (4, -4.1){};

\node[event-left] at (\leftMost,-5) {At time $\timeOnchainPropagation$ after sending $\pre$.};
\node[event-enter] at (1, -5){};

\draw[->] (1,-5.8) -- node[arrowLabel] {$(\msgFinalize, 2, \conId)$} (4,-5.8);

\node[event] at (4,-6.1) {
	If $\Validate(6; \manager^{\conId}) = \msgBad$, discard.\\
	Let $\teeSet_\msgOk \define \{\tee \in \manager^{\conId}.\conPool: (\cdots;\tee) \in \manager^{\conId}.\conWatchChalRes\}$.\\
	Set $\manager^{\conId}.\conPool \define \teeSet_\msgOk \cup \manager^{\conId}.\conPool[0]$,
	and $\manager^{\conId}.\conWatchChalBlock \define \bot$.
};
		\node[event-enter] at (4, -6.1){};

\node[event-left] at (\leftMost,-7.3) {$\res = \tee_E(\ledger, \manager^{\conId}.\conWatchChalRes)$};
		\node[event-enter] at (1, -7.3){};

\draw[<-] (0,-8.1) -- node[arrowLabel] {$(\res)$} (1,-8.1);
\end{tikzpicture}
	\end{center}
	\caption{Detailed watchdog challenge protocol.}
	\label{fig:diagram-challenge-watchdogs}
\end{figure}

Since the executor enclave $\enclave_E$ is dependent on the confirmation message from all watchdog enclaves, it is necessary to allow $\enclave_E$ to challenge the watchdog enclaves as well (Figure~\ref{fig:diagram-challenge-watchdogs}).
In this case, the executor enclave acts as the challenger and all watchdog enclaves need to provide a confirmation message as response.
At the end of this challenge-response protocol, all unresponsive watchdog enclaves are removed from the execution pool.
The executor enclave then continues performing the execution with all confirmations obtained during this procedure.
Again, $\manager$ only accepts responses if the watchdog executed the state update correctly, hence, ensuring that a watchdog either performs the correct state update or is kicked from the pool.

\subsection{Security Remarks}
\label{sec:protocol/security-remarks}

To keep the protocol description compact, we omitted some security features from the specification, which we explain in this section.

	Allowing unrestricted execution requests comes with the problem that malicious users can send requests whose execution takes a disproportional amount of time, e.g., due to infinite loops.
	If the execution time exceeded the boundaries defined by the on-chain timeouts, malicious users could exploit this behavior to kick honest operators from an execution pool.
This operator \emph{denial of service} attack harms the liveness property of the system.
In order to mitigate the vulnerability, we introduce an upper bound to the computation complexity of a single contract execution.
Once the bound is reached, the executor enclave stops executing and reverts the state but still provides a valid output.
The timeouts in the system are set such that an honest operator cannot be kicked from an execution pool even if an execution takes the maximum amount of computation.
The same applies to update and creation requests, where failed creations return a \textit{fail confirmation} that can be submitted to the manager instead of the creation confirmation.
A fail confirmation triggers the manager to mark the contract as crashed.
Note that the \sysname system still supports the execution of arbitrary complex smart contracts as the timeouts and hence the upper bounds can be set arbitrarily high (cf. Appendix~\ref{sec:high level protocol/timeouts}).
	Additionally, all contracts of an operator are executed and challenged independently, and thus, contracts do not block each other.

While we have assumed that all operators run only one \sysname enclave, multiple enclaves can be created in practice.
This enables the opportunity of a \emph{sybil attack}, where a malicious operator generates multiple \sysname enclaves to increase its share in the system and hence harm the liveness property.
This attack can be mitigated by forcing an operator to deposit funds at each enclave registration and which will be paid back to the operator only if she behaves honestly.
We note that this deposit is independent of any contract and its parties.
Now, such an attack is directly linked to financial loss.
See Section~\ref{sec:extensions} for more discussions about incentives and fees.

In order to enhance \emph{privacy}, neither users nor operators send inputs or respectively execution results in clear.
Instead, users encrypt inputs using hybrid encryption based on the public key of the executor enclave.
Additionally, users specify a symmetric key in their execution request, which is used to encrypt the result of the execution when sent back to the user.
This way, inputs and results are private and cannot be eavesdropped by a malicious operator.

The term \emph{griefing} denotes attacks where an adversary forces an honest party to interact with the blockchain in order to generate financial damage to this party.
Especially when blockchain transactions require high fees, such attacks pose serious vulnerabilities.
In regards to challenges within the \sysname protocol, we mitigate the attack surface for griefing attacks by incorporating a mechanism in the manager that fairly splits the fees for challenge and response between the challenger and the challenged party.
The same mechanism can be used for the contract creation process.

	An adversary executing a \emph{clogging} attack sends many transactions to the system to prevent honest users from issuing transactions.
	In the context of \sysname, an off-chain clogging attack results in honest clients making an on-chain challenge to ensure that their requests will be processed.
	Hence, a successful clogging attack has to be performed on-chain.
	For the on-chain challenge, our system inherits the vulnerabilities of the underlying blockchain.

\section{Extensions}
\label{sec:extensions}

We simplified some protocol steps in order to make the protocol description more compact and easier to understand.
We discuss the most important extensions and their benefits in this section.

\noindent\textbf{Contract \& Operator Lifecycle.}
A mechanism that releases enclaves from their execution duty can be integrated.
This allows operators to voluntarily withdraw their enclaves from an execution pool.
On the one hand, terminated contracts can be closed, which releases all pool enclaves from their execution duty.
On the other hand, it enables to withdraw a single enclave and exchanging it by a randomly chosen replacement enclave.
Additionally, a replacement strategy is also applicable to the scenarios in which enclaves are kicked.
The latter extension reduces the chance of a contract crash, the event in which no more operator remains.
We stress that these extensions can easily be achieved by adding the functionality to our \sysname program and the manager.
In case a contract is idle for a long time, an extension may be implemented that allows operators to \emph{hibernate} their respective enclave.
The enclave state can be stored on disk by encrypting it with a key that is kept alive in the hibernating enclave; thus, only requiring minimal overhead in memory.
The \sysname program ensures freshness by synchronizing with the blockchain; thus, preventing rollback attacks.

\noindent\textbf{Incentives.}
Although \sysname provides security not only against rational but also byzantine adversaries, it is beneficial to introduce incentives for operators to join the system and act honestly.
Moreover, operators can be compensated for on-chain transactions.
Such incentives can be achieved by introducing execution fees paid by the users to the operators.
We expect these fees to be significantly lower than Ethereum transaction fees since replication of computation is only required among a small pool.
Additionally, registration fees for operators can be used to mitigate the risk for sybil attacks.

	By mitigating these attacks and due to the random assignment of enclaves to contract pools, operators can only actively enforce centralization at high cost.

\noindent\textbf{Efficiency Improvements.}
Instead of propagating each contract invocation, a more fine-grained distinction based on the action can be added.
In particular, a simple state retrieval must not be propagated.
In order to improve the efficiency of the manager, messages and responses are not stored persistently.
Instead, only their hashes are stored and the actual data is propagated via events.
Moreover, the total on-chain transactions can be reduced by letting the executor enclave challenge only the unresponsive watchdog enclaves.

\section{Security Analysis}
\label{sec:security}

In this section, we present security considerations of \sysname based on the adversary model stated in Section~\ref{sec:adversary-model}.

\subsection{Protocol Security}
\label{sec:protocol-security}

We analyze the security of our protocol under the assumption of an IND-CPA secure encryption scheme, an EU-CMA secure signature scheme and a collision resistant hash function in the following.
We present definitions of correctness, \mbox{$\epsilon$-liveness} and state privacy.

\subsubsection{Correctness}
\label{app:protocol-security/correctness}

We define a state update as the evaluation of a transition function $f$, which receives as inputs a user $\user$, a user input $\move$ and a copy of the blockchain $\ledger$.
The \emph{correctness} property states that each state update evaluates the transition function as defined by the contract code with valid inputs, i.e., $\user$ is the (potentially malicious) client triggering the transition, $\move$ the input of $\user$ and $\ledger$ a valid copy of the blockchain that is at most $\tau^{\txt{off}}_{\txt{slack}}$ behind the main chain.

\begin{claim} [Correctness]
	The \sysname protocol satisfies correctness.
\end{claim}

We first note that according to our adversary model, a corrupted operator may delete any message intended for her enclave or generated from her enclave.
However, the correct execution of the \sysname program inside the enclave cannot be influenced.
When an operator creates a \sysname enclave, the registration process ensures that the new enclave indeed runs the \sysname program.
To this end, our protocol utilizes the TEE attestation mechanism, which generates a verifiable statement that the enclave is running a specific program.
Upon registration with the manager $\manager$, $\manager$ checks the validity of the attestation statement as well as the blockchain evidence, the signed hash and number of the latest block known to the enclave.
$\manager$ only registers the enclave in the system if the new enclave is running the \sysname program and is not further behind than maximally $\tau^{\txt{off}}_{\txt{slack}}$.
Finally, the TEE integrity and confidentiality guarantees ensure that a malicious operator cannot modify the enclave's code, tamper with its state or access its private data, in particular, its signature keys.

During the creation of a contract, the pool enclaves attest the code of the installed contract to the creation enclave.
The creator checks that the code is consistent with the hash stored in the manager before signing a creation confirmation.
Hence, it is not possible, without breaking the EU-CMA security of the signature scheme or the collision resistance of the hash function, to create a valid creation confirmation for a contract with different code than specified by the creation request.

Next, contract state updates can only be triggered by invoking the executor enclave with an execution request or invoking a watchdog enclave with an update request.
The correctness of the latter is reduced to the correctness of the former.
To see this, we observe that any update request to a watchdog enclave requires to be signed by the executor enclave.
Clearly, the executor enclave only signs updates corresponding to its own executions.
Therefore, an adversary cannot forge incorrect update request without breaking the unforgeability of the signature scheme.
Also, the executor enclave can only issue a new state update if all watchdogs confirmed the previous one.
Hence, it is not possible to tamper with the order in which the update requests are provided to a watchdog enclave.

As stated before, the TEE integrity guarantees ensure the correct execution of the program code and hence the correct execution of the smart contract.
It follows that a state update can only be achieved by providing inputs to the executor enclave.
The executor enclave receives a signed message containing the action $\move$ from user $\user$ and the relevant blockchain data from its operator.
In Section~\ref{sec:high level protocol/synchronization}, we describe how our protocol achieves secure synchronization between the executor enclave and the blockchain.
In particular, the synchronization mechanism ensures that the blockchain data accepted by an enclave is correct and complete in regard to a correct blockchain copy that is at most $\tau^{\txt{off}}_{\txt{slack}}$ behind the main chain.
This guarantees that $\ledger$, represented by the received blockchain data, is a synchronized copy of the current blockchain.
In order to protect inputs by honest users $\user$, $\move$ needs to be signed by $\user$.
This means an adversary cannot tamper with the input without breaking the signature scheme.

Finally, we note that each \sysname enclave maintains a list of received messages.
Since an honest user randomly selects a fresh nonce for each execution request, replay attacks can be detected and prevented by any executor enclave.

\subsubsection{Liveness}
\label{sec:protocol-security/liveness}

The liveness property states that every contract execution initiated by an honest user $\user$ will eventually be processed with high probability.
In case of a successful execution, a valid execution response is given by the executor.
Unsuccessful execution can only happen in case of a contract \emph{crash}.
In this event, the contract execution halts and neither honest nor malicious users can perform successful contract executions anymore.
We emphasize that the pool size can be set such that crashes happen only with negligible probability.
In particular, for $\epsilon$-liveness, the probability of a crash is bounded by $1-\epsilon$.

\begin{claim} [$\epsilon$-Liveness]
	Let $n$ be the total number of enclaves in the system, $m$ be the number of malicious operators' enclaves and $s$ be the contract pool size.
	The \sysname protocol satisfies $\epsilon$-liveness for $\epsilon = 1 - \Pi^{s-1}_{i = 0} (\frac{m-i}{n-i}) > 1 - (\frac{m}{n})^s$.
\end{claim}

Whenever user $\user$ sends an execution request to the executor enclave $\enclave_E$, $\user$ either directly receives a response or $\user$ challenges $\enclave_E$ via the manager $\manager$.
If $\enclave_E$ does not respond within some predefined timeout, it will be kicked out of the execution pool and one of the watchdog enclaves takes over the executor role.
User $\user$ can now trigger the execution again by interacting with the new executor enclave.
During the execution, the executor enclave $\enclave_E$ requires confirmations from all watchdog enclaves in order to produce a valid result.
However, watchdog enclaves cannot stall the execution forever, since $\enclave_E$ is able to challenge them via the manager as well.
All unresponsive watchdog enclaves will be kicked out of the execution pool and the confirmations from the remaining watchdogs are enough to create a result.
We stress that all timeouts are defined in Appendix~\ref{sec:high level protocol/timeouts} with great care to ensure that honest operators have enough time to respond.
For example, the timeout for the executor challenge is sufficient to allow the executor enclave to challenge the watchdog enclaves twice; once for a currently running off-chain execution and once for the challenged on-chain execution.

Although the protocol guarantees that honest operators' enclaves will never be kicked, there is a small probability that an execution pool consists only of malicious operators' enclaves.
If all enclaves are kicked out of the execution pool, the contract execution crashes.
Let $n$ be the number of total registered enclaves and $m$ denote the number of enclaves controlled by malicious operators.
The execution pool size is given by $s$.
The probability of a crash is equal to the probability that only malicious operators' enclaves are within an execution pool.
This is bounded by $\epsilon = 1 - \Pi^{s-1}_{i = 0} (\frac{m-i}{n-i}) > 1 - (\frac{m}{n})^s$.
Hence, \sysname achieves $\epsilon$-liveness.

Assuming a total of $n=100$ registered enclaves and $m=70$ of them are controlled by malicious operators.
Even in this setting with a large share of malicious operators, \sysname achieves liveness with $\epsilon>92\%$  for a pool size of just 7.
If only half of the operators are malicious, i.e., $m=50$, \sysname achieves liveness with $\epsilon>99\%$  for the same pool size of 7.
For $m =10$ malicious operators, a pool size of only 3 yields a liveness with $\epsilon > 99 \%$.
	For the same scenario of 10\% malicious operators and assuming 40 millions contracts running in \sysname, the pool size of 11 results in a probability of more than 99\% that there is no crash at all in the whole system.
	See Fig.~\ref{fig:probs} for an illustration of the probability of no crashes depending on the number of contracts for different pool sizes.

\subsubsection{State Privacy}

The \emph{state privacy} property says that the adversary cannot obtain additional information about a contract state besides what she learns from the results of contract executions alone.

\begin{claim} [State Privacy]
	The \sysname protocol satisfies state privacy.
\end{claim}

The smart contract's state is maintained by the enclaves within the execution pool.
According to our adversary model (see Section~\ref{sec:adversary-model}), the TEE provides confidentiality guarantees.
In particular, the execution of an enclave does not leak any data.
Hence, the smart contract's state is hidden from the adversary, even if the enclave's operator is corrupted.
The only point in time during the \sysname protocol when information about the contract's state is revealed is at the end of the execution protocol.
However, the data provided as a result contains only public state and hence does not reveal anything about the private state.
During the execution protocol, the executor enclave propagates the new state to all watchdog enclaves.
However, the transferred data is encrypted using an IND-CPA secure encryption scheme.
The security of the scheme guarantees that an adversary seeing the message cannot extract information from it.

While an enclave only publishes outputs after successful executions, we need to show that each produced output is final.
In particular, a succeeding executor must not be able to revert to a state in which a published output should not have been produced.
To this end, the state of the executor enclave producing a particular output needs to be replicated among all other enclaves before revealing the actual output.
This property is achieved by the state propagation mechanism in our protocol.
An enclave only returns an output if all enclaves in the pool confirm the corresponding state update.
The EU-CMA secure signature scheme guarantees unforgeability of the confirmations.
Hence, each confirmation guarantees that the corresponding enclave has updated its state correctly.
Further, the correctness property of our protocol (cf. Section~\ref{app:protocol-security/correctness}) ensures that an enclave is always executed with a correct blockchain copy, and hence, is always aware of the correct pool composition.
This means that an output can only be returned if the whole pool has received the corresponding state update.

\subsection{Architectural Security}
\label{sec:architectural-security}

We further examine the architectural security of enclaves.
The case of a user or TEE operator going offline by turning off their machine is covered in the protocol security (cf. Section~\ref{sec:protocol-security}); here we focus on parties that follow the protocol, trying to gain an unfair advantage in various ways.

The adversary might try to perform a memory corruption attack on the client used by users to interact with the executor (e.g., to send inputs).
To mitigate this risk, the software should be implemented in a memory-safe language, like Python or Rust, and be open source so that it can be easily inspected.

A malicious TEE operator can also try mounting a memory-corruption or a side-channel attack on its TEE.
As mentioned in~\ref{itm:TEE-i-c}, we assume that the TEE protects the confidentiality of the enclave and prevents leakage.
However, in practice, cache-based side-channel attacks have been successfully demonstrated also on ARM processors~\cite{lipp2016armageddon}.
While we want to stress that our ARM TrustZone-based implementation is a research prototype and the design is TEE-agnostic, the risk of these attacks can be mitigated by making the TEE opt-out of shared caches and flush private caches upon context switch, as proposed in~\cite{sanctuary}.
Alternatively, a more advanced TEE design can be used~\cite{sanctum,sanctuary,cure}.
Moreover, if the enclave code has an exploitable memory-corruption vulnerability, it is possible to mount a memory-corruption attack against it.
One way to mitigate this risk, and hence, realize our assumption~\ref{itm:TEE-no-exploit}, is to use a memory-safe language for our smart contracts (in our case, Lua), or to deploy a run-time mitigation (like CFI~\cite{abadi2005cfi}).
Yet, in practice, an adversary might still be able to compromise an enclave. In this case, only the contracts of this enclave are affected. The consequences depend on the role of the enclave: for an executor enclave, the adversary gets full control over the contract; for a watchdog enclave, the adversary can only break state privacy.

Finally, an adversary might build a malicious smart contract with the goal of compromising secrets owned by other contracts or blocking an enclave by entering into an infinite loop.
We mitigate against the first scenario by ensuring that only one smart contract is executing at any given time in an enclave, so that no foreign plain text secrets are present in memory at any point during contract execution.
In case of multiple enclaves running on the same system, the TEE is isolating enclaves from each other such that no contract can tamper with another (cf. assumption~\ref{itm:TEE-i-c}).
To handle infinite loops, we leverage a Lua sandbox~\cite{sandbox.lua}, which interrupts the execution of the Lua code after a predetermined number of instructions has been issued and disables access to unsafe functions and modules.

\section{Implementation}
\label{sec:impl}
In order to evaluate \sysname, we implemented a prototype for the manager and the enclaves, which uses TrustZone for the enclaves themselves and Lua as the smart contract programming language. We open source our prototype implementation to foster future research in this area\footnote{\mbox{\url{https://github.com/AppliedCryptoGroup/PoseCode}}}.
We describe each of them in the following.

\noindent\textbf{Manager.}
For the manager we use an Ethereum smart contract written in Solidity, which we will refer to as \emph{manager} in the following.
Even if this implementation is based on Ethereum, we note that our design can be realized on any blockchain supporting rich smart contracts.
The manager keeps a list of all registered enclaves in the network as well as a list of all deployed contracts, including their public information, e.g., the address of the current executor.
As mentioned in the protocol described in \Cref{sec:protocol/description}, the manager provides functions to register an enclave, create a new \sysname contract, deposit or withdraw money, and functions to challenge the current executor or any of the watchdogs.
To synchronize all participants, every time a challenge related function was called it will throw an appropriate Solidity event.

\noindent\textbf{Enclaves.}
The contract creator, executor, and watchdogs are enclaves running in a \ac{tee}.
As our protocol is TEE-agnostic and all commercial TEEs exceed smart contracts' on-chain requirements on memory/computational-power capabilities significantly, we chose to use ARM TrustZone~\cite{trustzone} for our prototype. TrustZone features a traditional programming model (OS, and user-space applications with standard library), and the Open Portable Trusted Execution Environment (OP-TEE) OS~\cite{optee} already supports a large fraction of standard functionality, and hence, does not force us to reimplement this for the contract execution environment. 
TrustZone supports two execution modes: secure world and normal world.
The system's memory can be freely distributed among these worlds.
The secure world is an trusted OS which is completely independent from the normal OS, which in our case is Linux.
Code running in the secure world is called a \emph{Trusted App} (TA).
A TA may only communicate with the normal world via shared memory regions, which are explicitly allocated as such.
We implement the \sysname enclaves as TAs.
Computations in the secure world have native performance; yet, switching between worlds has a constant but negligible overhead (in our tests around 449µs).
TrustZone does not impose memory limits for secure world.
While we leverage the traditional TrustZone concept, recent versions add support for a S-EL2 hypervisor to allow multiple strongly isolated enclaves that allows \sysname to scale better on these platforms.
Most basic cryptographic functions are provided by the OP-TEE TA library, such as AES and TLS.
Note that TrustZone itself does not standardize a remote attestation implementation itself, but industry~\cite{samsung-attestation, qualcomm-attestation, google-attestation} and OP-TEE implementations exist\footnote{\url{https://github.com/OP-TEE/optee_os/pull/5025}}.
Remote attestation can also be used to prove a certain set of software defenses is active in the enclave.
In our prototype, we leveraged OP-TEE's remote attestation functionality to attest the enclave after setting up the runtime. To leverage this feature, the \sysname enclave requests a signed attestation report from the attestation PTA (Pseudo Trusted App), essentially a kernel module of the OP-TEE OS in secure world. The keys for signing the attestation report are derived using hardware device information and stored persistently after generation (using Secure Storage, or "Trusted Storage", as defined by GlobalPlatform’s TEE Internal Core API specification).

To properly interact with the Ethereum-based manager, we also adapted and deployed an Ethereum wallet for embedded devices~\cite{walletrepo}, enabling the enclaves to create ECDSA signatures, Keccak hashes, handle encoding, and create transactions to call the manager.
For \sysname contracts, we use the scripting language Lua~\cite{lua}.
It is a well-established, fast, powerful, yet simple language written in C.
Lua as well as the enclave itself allow arbitrary computation.
We ported the Lua interpreter to run inside the TA, by stripping out operations unsupported by the TA, such as file access.
After each execution step, the enclave returns to the normal world while keeping the contract's Lua session alive.
When the normal world receives an input from a user, it invokes the TA with these inputs to continue the Lua execution. 
To update the enclave runtime, different approaches are possible in practice, e.g., the manager could announce an update and all outdated enclaves would shut themselves down after a timeout. Honest operators then would incrementally trigger an enclave replacement during the timeout period.

\section{Evaluation}
\label{sec:eval}
This section examines \sysname regarding complexity and performance. In the following, we will report absolute performance numbers and discuss these in relation to Ethereum itself, but also compare to existing works based on TEEs, namely FastKitten and Bitcontracts. FastKitten has a highly similar set of tested smart contracts, so a comparison can put our numbers in perspective. For Bitcontracts, we reimplemented Quicksort with the same experiment setup. Note, that the smart contracts can still be implemented differently, and the performance and the TEE differ.

\noindent\textbf{Complexity.}
Running a \sysname contract in the benign case, i.e., if all involved enclaves respond, requires exactly two blockchain interactions for the setup.
Each user of a contract also needs one blockchain interaction each time the user deposits or withdraws money regarding the contract.
However, as \sysname does not require a fixed collateral for the setup, the money transactions do not inherently prevent the contract from execution---except the specific contract demands it.
Otherwise, when either the executor or any watchdog fails to respond, each challenge requires two blockchain interactions.
The delay incurred by our challenge protocol is dominated by the on-chain transactions.
	This holds also for other off-chain solutions, e.g., state-channels~\cite{Sprites,PerunStateChannels,L4}, Plasma~\cite{Plasma,khalil2018commit}, Rollups~\cite{arbitrumRollup,optimisitcRollup} and FastKitten~\cite{fastkitten}.
	For instance, the time it takes for an honest executor to kick a watchdog is $325 s$ on average. We discuss timeout parameters and the challenge delay more thoroughly in~\Cref{sec:high level protocol/timeouts}.
In the worst-case, a malicious operator does not respond to the off-chain messages but to the challenges in every execution step, which would effectively reduce \sysname's execution speed beneath that of the blockchain.
However, such an attack requires continuous blockchain interactions from the malicious party and hence entails costs for every execution step (cf. \Cref{sec:eval_manager} ``Manager'').

\noindent\textbf{Test Setup.}
We deployed a test setup with our prototype implementation for performance measurements.
The test setup consists of five devices.
For the enclaves we deployed three Raspberry Pi 3B+ with four cores running at 1.4GHz.
These are widely available and cheap devices that support ARM TrustZone.
As state updates are small (just the delta to the previous state) and watchdogs receive and process the state updates in parallel, we do not expect an increase of the pool size to significantly influence the evaluation.
Further, we used \texttt{ganache-cli} (6.10.2) to emulate a Ethereum blockchain in our local network, which runs the Solidity contract that implements the manager.
Finally, a fifth device emulates multiple users by simply sending out network requests to both the manager and enclave operators, which are all connected via Ethernet LAN.

\noindent\textbf{Manager.}
\label{sec:eval_manager}
As the \sysname manager is implemented as an Ethereum smart contract, interactions with it incur some costs in the form of Gas.
The costs of all implemented methods of the Solidity contract are listed in \Cref{fig:gas-costs}.
The first five methods are used for benign \sysname contract execution.
The second part of the table shows methods that are required for challenges, including the response and timeout methods to resolve them.
In terms of storage, each additionally registered enclave will require 64~bytes and each contract 288~bytes + (pool~size~$\times$~32~bytes) of on-chain storage.

\begin{table}[t]
	\small
	
	\caption{Cost of executing the methods of the \sysname manager.
		The USD costs were estimated based on the prices (Gas to GWei and ETH to USD) on May. 8, 2022~\cite{gasprice,ethprice}. *For comparison, these are the costs of popular operations on Ethereum.
	}
	\centering
	
	\begin{tabular*}{\columnwidth}{!{\extracolsep{\fill}}>{\ttfamily}lrr}
		\toprule
		\multirow{2}{*}{\rmfamily\textbf{Method}} & \multicolumn{2}{c}{\textbf{Cost}} \\
		& \multicolumn{1}{c}{\textbf{Gas}} & \multicolumn{1}{c}{\textbf{USD}} \\
		\midrule
		registerEnclave
		& 175\,910 & 13.23 \\
		initCreation               & 198\,436 & 14.91 \\
		finalizeCreation           &  79\,545 & 5.98 \\
		deposit                    &  37\,255 & 2.80 \\
		withdraw                   &  36\,997 & 2.78 \\ 
		\midrule
		challengeExecutor          &  54\,654 & 4.11 \\
		executorResponse           &  51\,478 & 3.87 \\
		executorTimeout            &  53\,327 & 4.01 \\
		challangeWatchdogsCreation & 231\,286 & 17.38 \\
		challengeWatchdog          & 131\,362 & 9.87 \\
		watchdogResponse           &  36\,257 & 2.72 \\
		watchdogTimeout            &  52\,142 & 3.92 \\
		\midrule\\[-1.5em]\midrule
		\textrm{simple Ether transfer*} &  21\,000 & 1.58 \\
		\textrm{create CryptoKitty*}    & 250\,000 & 18.78 \\
		\bottomrule
	\end{tabular*}

	\vspace{1ex}
	\label{fig:gas-costs}
\end{table}

\noindent\textbf{Contract Execution.}
To measure and demonstrate the efficiency of \sysname contract execution, we implemented three applications as Lua code in our test setup. 
All time measurements are averaged over 100 runs.
Regardless of the used contract, setting up an executor or watchdog enclave with a Lua contract takes 189ms. Creating an attestation report for the enclave takes another 367ms with OP-TEE's built-in remote attestation using a one-line dummy contract. For our biggest contract, Poker, the attestation takes 377ms, resulting in a total setup time of 566ms. In contrast, FastKitten needs 2s for enclave setup. Note that FastKitten needs an additional blockchain interaction.
Multiple contracts run by a single operator are executed in parallel, including network communication. Thus, the number of enclaves, contracts and transactions a single operator can process depends on the operator's hardware. As modern servers CPUs feature 128 cores~\cite{ampere_altra_max}, and servers often feature multiple CPUs, we do not expect parallel execution to affect performance significantly. However, to prevent overload, the number of pools an operator participates in can be limited. %

\noindent\textit{Rock paper scissors.}
This is an implementation of the popular game with two players.
Unlike traditional smart contracts, we can leverage \sysname's private state to simply store each player's input, instead of having to use much more complex multi-round commitments.
The resulting smart contract is 27 lines of code (LoC).
Disregarding the delay caused by human players, the execution time of one round with two user inputs is 32ms. In comparison, FastKitten only needs 12ms, but is also running on a much more powerful machine.
In contrast, executing this game on Ethereum would take around 5 minutes for each round (20 confirmation blocks, 15s block time each).

\noindent\textit{Poker.}
We have also implemented Poker as a multi-party contract running over multiple rounds.
Note that in \sysname, the poker game can be implemented as an ongoing cash game table, i.e., players may join or leave the table at any time, as contracts in \sysname do not have to be finite.
Each round consists of three phases each requiring an input from all users.
The resulting smart contract is 209 lines of code (LoC).
We execute the contract with five players who have their deposit ready at the start, with a total execution time of 199ms (vs. 45ms in FastKitten, but again, on a more powerful machine).
Playing this game on Ethereum would take 5 minutes per player input.

\noindent\textit{Federated Machine Learning.}
For this application, users can submit locally trained models, which will be aggregated to a single model by the contract.
Any user can then request the new model from the contract.
For our measurements, each user trained a convolutional neural network consisting of $431\,080$ individual weights on the MNIST handwritten digits dataset~\cite{mnist}.
For aggregation, the contract averages every existing weight with the corresponding weight sent by the user.
The smart contract itself is only 5 LoCs, as we load the existing weights separately.
Each aggregation took 238ms, which demonstrates the efficiency of \sysname.
Trying to execute the same function on Ethereum, for each aggregation, storage of the weights alone would exceed 1 billion gas (assuming 4 bytes float per weight) and the calculation over 3.4 million gas (8 gas per weight).

\noindent\textit{Quicksort.}
We have also implemented Quicksort to sort a hardcoded input array of 2048 random integers, as done in Bitcontracts~\cite{bitcontracts}.
The resulting smart contract is 29 lines of code (LoC).
The total execution time of the contract is 20ms. Compared to the 6ms in Bitcontracts, we use a less powerful machine (Bitcontracts uses an AWS T2.micro instance with a recent Intel processor at 3.3Ghz), while our performance measurement also includes additional steps like context switches and the setup of the enclave runtime.
Executing this Quicksort contract on Ethereum would cost around 6.5 million gas.

\noindent\textbf{Watchdog State Updates.}
When an executor operator has been dropped, a watchdog takes over execution. For this to work, state changes are distributed to the watchdogs. Storing the current state and restoring it on a watchdog takes 17ms for the poker contract (averaged over 100 runs, corrected for network latency), which also has the biggest state among the ones we implemented.

\noindent\textbf{Enclave Teardown.}
After an executor enclave is not expecting further inputs and finished the smart contract execution, the execution environment has to be cleaned up for the next smart contract, i.e., cryptographic secrets and the smart contract in the shared memory need to be zeroed. This takes 25ms.

\begin{table}[t]
\small
\definecolor{crossRed}{RGB}{192, 0, 0}
\definecolor{checkGreen}{RGB}{112, 173, 71}
\newcommand{\tablecross}{\textcolor{crossRed}{\ding{55}}}
\newcommand{\tablecheck}{\textcolor{checkGreen}{\ding{51}}}
\newcommand{\tablemid}{\normalsize\bfseries\textasciitilde}
\newcommand{\tableka}{?}

\caption{Overview of related work, $n$ denotes the number of transactions.}

\begin{tabular*}{\columnwidth}{!{\extracolsep{\fill}}lcccc}
\toprule
& \rotatebox{90}{No collateral}
& \rotatebox{90}{Private state}
& \rotatebox{90}{\parbox{\widthof{(optimistically)}}{Blockchain \\[-.3ex]interactions\\[-.3ex](optimistically)}}
& \rotatebox{90}{\parbox{\widthof{No regular}}{Non-fixed\\[-.3ex]lifetime\\[-.3ex]\& group}}
\\
\midrule
Ethereum~\cite{ethereum}
& \tablecheck   & \tablecross & \textcolor{crossRed}{$O(n)$} & \tablecheck \\
MPC~\cite{kumaresan2015use,KumaresanVV16,KumaresanB16}
& \tablecross & \tablecheck & \textcolor{checkGreen}{$O(1)$} & \tablecross \\
State Channels~\cite{Sprites,PerunStateChannels,L4}
& \tablecross & \tablecross   & \textcolor{checkGreen}{$O(1)$} & \tablecross \\
VM-based~\cite{Arbitrum,ace,bitcontracts}
& \tablecross    & \tablecross &  \textcolor{crossRed}{$O(n)$}    & \tablecheck    \\
\midrule
Ekiden~\cite{ekiden}
& \tablecross & \tablecheck & \textcolor{crossRed}{$O(n)$} & \tablecross   \\
FastKitten~\cite{fastkitten}
& \tablecross & \tablecheck & \textcolor{checkGreen}{$O(1)$} & \tablecross \\
\midrule
\sysname       
& \tablecheck & \tablecheck & \textcolor{checkGreen}{$O(1)$} & \tablecheck \\
\bottomrule
\end{tabular*}
\label{tab:rw}
\end{table}

\section{Related Work}
\label{sec:relatedwork}

Ethereum~\cite{ethereum} is the most prominent decentralized cryptocurrency with support for smart contract execution.
However, it is suffering from very high transaction costs and data used by smart contracts is inherently public.

Hawk~\cite{kosba2016hawk} aims for improving the privacy by automatically creating a cryptographic protocol from a high-level program in order to allow computation on private data without disclosing it.
However, this complex cryptographic layer further decreases performance of the system and increases costs.
Similarly, approaches based on \ac{mpc}~\cite{kumaresan2015use,KumaresanVV16,KumaresanB16} distribute the computation between multiple parties such that no party can access the cleartext data.
These approaches have substantial overhead in performance, communication and collateral required.

One approach to alleviate the complexity limitation are state channels~\cite{Sprites,PerunStateChannels,L4}, which enable parties to lock some funds on the blockchain, execute complex contracts off-chain, and finally commit the results of the contract to the blockchain.
This is efficient if all parties agree on the results; otherwise, the dispute can be solved on-chain, which takes longer and is more expensive.

Arbitrum~\cite{Arbitrum} represents a smart contract as a virtual machine (VM), which is executed privately by a number of ``managers''.
After execution, if all managers agree on the result of the computation, this result can be simply signed and committed to the blockchain, without the need to perform the computation on chain.
In case managers disagree, a bisection algorithm is used to compare subsets of the execution on chain and find which is the first instruction on which the managers disagree, then punish the malicious manager(s).
Hence, as long as at least one manager is honest, the correct result is computed.
While computationally efficient, this on-chain protocol is still relatively expensive, so Arbitrum also includes financial incentives to encourage the managers to behave.
The managers have full access to the data used by the VM, so confidentiality is broken if even one manager is malicious.
Note that unlike Arbitrum, \sysname does not require multiple parties to execute the smart contract: the watchdog enclaves just need to acknowledge the new states, unless the executor enclave fails.

ACE~\cite{ace} and Bitcontracts~\cite{bitcontracts} are similar to Arbitrum, but they allow the results of contract executions to be approved by a configurable quorum of service providers, not necessarily all of them.
Unlike \sysname, ACE does not support private state and requires on-chain communication per contract invocation.
Although the transaction is computed off-chain, the invocation and the result are registered on-chain.
Further, Arbitrum and ACE require changes to the blockchain infrastructure, hence, they are harder to deploy in practice.

Ekiden~\cite{ekiden} is also an off-chain execution system that leverages TEE-enabled \emph{compute nodes} to perform computation and regular \emph{consensus nodes} that interact with a blockchain.
The major drawback of Ekiden is that it requires every computation step to retrieve its initial status from the blockchain, and it only supports input from one client at a time.
Moreover, the atomic delivery of the output of each step requires to wait for publication of the updated state before the output is made available to the client.
Hence, any highly interactive protocol with multiple participants (like a card game, for instance) would incur significant delays between turns just to wait for the blockchain.
The paper evaluates Ekiden on a fast blockchain, Tendermint, but it does not quantify its latency for interactive protocols on mainstream blockchains like Ethereum or Bitcoin.
The Oasis Network uses an updated version of Ekiden~\cite{oasis}; yet, this version still requires to store state on the blockchain after each call.

FastKitten~\cite{fastkitten} also leverages TEEs to perform off-chain computation.
It assumes a rational attacker model, with financial incentives to convince all participants to follow the protocols.
If they all do, the communication happens directly between the TEE and them, thus dispensing with the high latency due to blockchain roundtrips.
However, FastKitten only supports contracts with a predefined list of participants and a limited lifespan.
It also requires the TEE operator to deposit as much as every participant combined as collateral.
\sysname lifts those restrictions: it enables long-lived smart contracts with an unknown set of participants and requires no collateral from the TEE owners.
Further, \sysname achieves strong liveness guarantees in the presence of byzantine adversaries, while FastKitten assumes a rational adversary.

ROTE~\cite{rote} is a system that detects rollback attacks on \acp{tee} by storing a counter on a number of other \acp{tee}.
This approach is similar to the watchdog enclaves used in \sysname to ensure that execution of a smart contract continues.
However, unlike \sysname, ROTE can only detect rollback attacks, but cannot prevent malicious operators from withholding the state.
SlimChain~\cite{slimchain} primarily aims at reducing on-chain storage, while still requiring blockchain interactions to store state commitments. Further, the paper does not address storage nodes crashing, which would lead to a liveness violation.
Pointproofs~\cite{pointproofs} proposes a new vector commitment scheme to reduce the storage requirements on blockchain validators. Although validators do not need to store all values of a smart contract, once a transaction provides these values, the execution is still performed on-chain.
However, \sysname works entirely off-chain in the optimistic case and ensures liveness with its protocol.

Chainspace~\cite{chainspace} proposes an entirely new distributed ledger platform focusing on sharding combined with a directed acyclic graph structure, while POSE extends established blockchains (e.g., Ethereum).
ResilientDB~\cite{resilientdb} proposes a consensus protocol that clusters validators' geo-location to minimize network overheads. 
In contrast, \sysname is a off-chain execution protocol for smart contracts.
Hyperledger Fabric Private Chaincode~\cite{privatechaincode} requires trust in handling the encryption key by the client or an \emph{admin}; thus, we deem it not applicable to permissionless blockchains, targeted by \sysname. %
Hyperledger \emph{Private Data Objects}~\cite{pdo}, an alternative to Private Chaincode, requires periodic blockchain interactions to store the state on-chain. 
This slows execution on contract calls to the speed of the blockchain, unlike \sysname, which executes contracts entirely off-chain in the optimistic case.
Hyperledger \emph{Avalon}~\cite{avalon} can outsource workloads to TEE enclaves. However, these workloads have to be self-contained, and thus, interactions by participants still require on-chain transactions, while \sysname can run interactive contracts completely off-chain (e.g., Poker).

\section{Conclusion}
Smart contracts have become an indispensable tool in the era of blockchains; yet, current approaches suffer from various shortcomings.
In this paper, we introduce \sysname, a novel off-chain execution protocol that addresses all of these shortcomings to enable much more versatile smart contracts.
We showed \sysname's security and demonstrated its feasibility with a prototype implementation.

\section*{Acknowledgements}
This work was supported by the European Space Operations Centre with the Networking/Partnering Initiative, the German Federal Ministry of Education and Research within the StartUpSecure funding program \textit{Sanctuary} (16KIS1417) and within the \textit{iBlockchain project} (16KIS0902), by the Deutsche Forschungsgemeinschaft (DFG, German Research Foundation) \textit{SFB 1119 – 236615297 (CROSSING Project S7)}, by the European Union’s Horizon 2020 Research and Innovation program under Grant Agreement No. 952697 (ASSURED), by the BMBF and the Hessian Ministry of Higher Education, Research, Science and the Arts within their joint support of the \textit{National Research Center for Applied Cybersecurity ATHENE}.

\begin{small}
\bibliographystyle{plain}
\bibliography{bibliography}
\end{small}

\appendix
\subsection{Supported Contracts}
\label{app: supported contracts}
The \sysname system supports contracts with a dynamic set of users of arbitrary size and an unrestricted lifetime.
The timeouts need to be set reasonable with respect to the expected execution time of the contracts to allow the execution of complex contracts and to prevent denial of service attacks at the same time.

Interaction between \sysname contracts can be realized by letting the TEE of the calling contract instruct its operator to request an execution of the second contract via the respective executive operator and wait for the response.
We deem the exact specification, e.g., enforce an upper bound on (potentially recursive) external calls to guarantee timely request termination, an engineering effort.
Calls from POSE contracts to on-chain contracts can be supported similarly to our payout concept (Appendix \ref{sec:protocol/payments}).

\subsection{Further Protocol Blocks}
\label{sec:app-further-protocol}

To keep the specification of the \sysname protocol in the main body simple and compact, we have excluded the formal specification of the creation process and the validation algorithms.
In this section, we present the full specification of these protocol parts.

\subsubsection{Creation protocol and creation challenges}
\label{sec:app-further-protocol/creation}
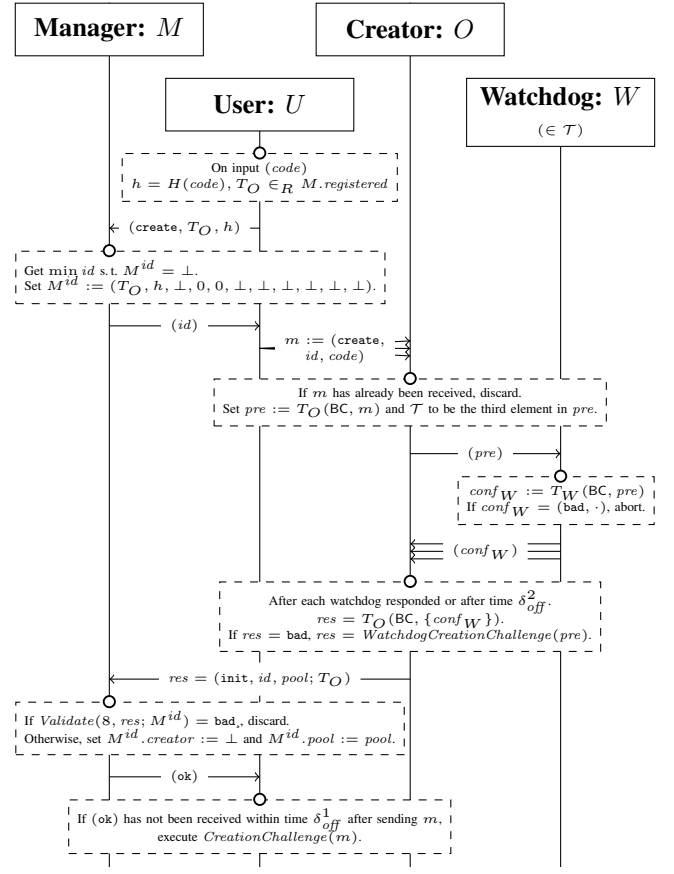
\begin{figure}
	\begin{center}
		\begin{tikzpicture}
		
		\newcommand{\rightMost}{8.25};
		\newcommand{\leftMost}{-0.25};
		
		\tikzset{header/.style={draw, font=\bf, fill = white, below, align=center, minimum width=2.5cm,minimum height=0.7cm}}
		\tikzset{event/.style={draw, font=\tiny, fill=white, dashed, below, align=center, minimum width=2cm,minimum height=0.5cm}}
		\tikzset{arrowLabel/.style={draw, draw = white, align=center, fill=white,font=\tiny}}
		\tikzset{event-right/.style={draw, font=\tiny, fill=white, dashed, below left, align=right, minimum width=2cm,minimum height=0.5cm}}
		\tikzset{event-left/.style={draw, font=\tiny, fill=white, dashed, below right, align=left, minimum width=2cm,minimum height=0.5cm}}
		\tikzset{event-enter/.style={draw, circle, line width=0.25mm, fill=white,inner sep=1.5pt}}
		
		\draw (1,-9.8) -- (1,1.7);
		\draw (3,-9.8) -- (3,0);
		\draw (5,-9.8) -- (5,1.7);
		\draw (7,-9.8) -- (7,0);
		
		\node[header] (U) at (1,1.7) {Manager: $\manager$};
		\node[header] (U) at (3,0.7) {User: $\user$};
		\node[header] (EO) at (5, 1.7) {Creator: $O$};
		\node[header] (WO) at (7,0.7) {Watchdog: $W$\\
			\tiny{$(\in \pool)$}};
		
		\node[event] at (3,-0.3) {On input $(\code)$\\
			$h = \Hash(\code), \tee_O \in_R \manager.\tees$};
		\node[event-enter] at (3, -0.3){};
		
		\draw[<-] (1,-1.3) -- node[arrowLabel] {$(\msgCreate, \tee_O, h)$} (3,-1.3);
		
		\node[event-left] at (\leftMost,-1.6) {Get $\min \conId$ s.\,t. $\manager^\conId = \bot$.\\ Set $\manager^\conId \define (\tee_O, h, \bot, 0, 0, \bot, \bot, \bot, \bot, \bot, \bot)$.};
		\node[event-enter] at (1, -1.6){};
		
		\draw[->] (1,-2.6) -- node[arrowLabel] {$(\conId)$} (3,-2.6);
		
		\draw[->] (3,-2.9) -- node[arrowLabel] {} (5,-2.8);
		\draw[->] (3,-2.9) -- node[arrowLabel] {} (5,-3.0);
		\draw[->] (3,-2.9) -- node[arrowLabel] {$m \define (\msgCreate,$\\$ \conId, \code)$} (5,-2.9);
		
		\node[event] at (5,-3.3) {
			If $m$ has already been received, discard.\\
			Set $\pre \define \tee_O(\ledger,m)$ and $\pool$ to be the third element in $\pre$.};
		\node[event-enter] at (5, -3.3){};
		
		\draw[->] (5,-4.3) -- node[arrowLabel] {$(\pre)$} (7,-4.3);
		
		\node[event-right] at (\rightMost,-4.6) {$\conf_W \define \tee_W(\ledger, \pre)$\\
			If $\conf_W = (\msgBad, \cdot)$, abort.};
		\node[event-enter] at (7, -4.6){};
		
		\draw[<-] (5,-5.5) -- node[arrowLabel] {} (7,-5.5);
		\draw[<-] (5,-5.7) -- node[arrowLabel] {} (7,-5.7);
		\draw[<-] (5,-5.6) -- node[arrowLabel] {$(\conf_W)$} (7,-5.6);
		
		\node[event] at (5,-6) {After each watchdog responded or after time $\timeOffchainPropagation$.\\
			$\res = \tee_O(\ledger, \{\conf_W\})$.\\
			If $\res = \msgBad$,
			$\res = \protWatchdogCreationChallenge(\pre)$.
		};
		\node[event-enter] at (5, -6){};
		
		\draw[->] (5,-7.3) -- node[arrowLabel] {$\res = (\msgInit, \conId, \conPool; \tee_O)$} (1,-7.3);
		
		\node[event-left] at (\leftMost,-7.6) {If $\Validate(8, \res; \manager^{\conId}) = \msgBad$¸, discard.\\
			Otherwise, set $\manager^\conId.\conCreator \define \bot$ and $\manager^\conId.\conPool \define \conPool$.};
		\node[event-enter] at (1, -7.6){};
		
		\draw[->] (1,-8.6) -- node[arrowLabel] {$(\msgOk)$} (3,-8.6);
		
		\node[event] at (3,-8.9) {If $(\msgOk)$ has not been received within time $\timeOffchainCreation$ after sending $m$,\\
			execute $\protCreationChallenge(m)$.};
		\node[event-enter] at (3, -8.9){};
		
		\end{tikzpicture}
	\end{center}
	\caption{Detailed creation protocol.}
	\label{fig:diagram-creation}
\end{figure}

\begin{figure}
	\begin{center}
		\begin{tikzpicture}
		
		\newcommand{\rightMost}{8.25};
		\newcommand{\leftMost}{-0.25};
		
		\tikzset{header/.style={draw, font=\bf, above, align=center, minimum width=2.5cm,minimum height=0.7cm}}
		\tikzset{event/.style={draw, font=\tiny, fill=white, dashed, below, align=center, minimum width=2cm,minimum height=0.5cm}}
\tikzset{arrowLabel/.style={draw, draw = white, align=center, fill=white,font=\tiny}}
\tikzset{event-right/.style={draw, font=\tiny, fill=white, dashed, below left, align=right, minimum width=2cm,minimum height=0.5cm}}
\tikzset{event-left/.style={draw, font=\tiny, fill=white, dashed, below right, align=left, minimum width=2cm,minimum height=0.5cm}}
\tikzset{event-enter/.style={draw, circle, line width=0.25mm, fill=white,inner sep=1.5pt}}
		
		\node[header] at (1,0) {User: $\user$};
		\node[header] at (4,0) {Manager: $\manager$};
		\node[header] at (7,0) {Creator: $O$};
		
		\draw (1,-7.8) -- (1,0);
		\draw (4, -7.8) -- (4,0);
		\draw (7,-7.8) -- (7,0);
		
		\draw[->] (0,-0.4) -- node[arrowLabel] {$(m)$} (1,-0.4);
		\draw[->] (1,-0.7) -- node[arrowLabel] {$m \define (\msgCreate, \conId, \code)$} (4,-0.7);
		
		\node[event] at (4,-1) {If $\Validate(7,m;\manager^\conId, \code) = \msgBad$, discard.\\
			Otherwise, set $\manager^\conId.\conExecChalMsg \define m$, $\manager^\conId.\conExecChalBlock \define \ledger.\latestBock$ and $\manager^\conId.\conExecChalRes \define \bot$.};
		\node[event-enter] at (4, -1){};
		
		\draw[->] (4,-2.1) -- node[arrowLabel] {$(m)$} (7,-2.1);
		
		\node[event-right] at (\rightMost,-2.4) {Handle $m$ like a message directly received by a user\\
			and receive $\res = (\msgOk, \cdot)$ from $\tee_O$.};
		\node[event-enter] at (7, -2.4){};
		
		\draw[->] (7,-3.5) -- node[arrowLabel] {$\txt{res} = (\msgInit,$\\$\conId,\conPool; \tee_O)$} (4,-3.5);
		
		\node[event] at (4,-3.8) {
			If $\Validate(8, \res;\manager^\conId) = \msgBad$, discard.\\
			Otherwise, set $\manager^\conId.\conCreator \define \bot, \manager^\conId.\conExecChalMsg \define \bot$,\\ $\manager^\conId.\conExecChalBlock \define \bot$ and $\manager^\conId.\conPool \define \conPool$.
		};
	\node[event-enter] at (4, -3.8){};
		
		\draw[->] (4,-5.3) -- node[arrowLabel] {$(\msgOk)$} (1,-5.3);
		
		\node[event-left] at (\leftMost,-5.6) {If $\res$ is not received within time $\timeOnchainExecution$ after sending $m$.};
		\node[event-enter] at (1, -5.6){};
		
		\draw[->] (1,-6.5) -- node[arrowLabel] {$(\msgFinalize, 1, \conId)$} (4,-6.5);
		
		\node[event] at (4,-6.8) {
			If $\Validate(9;\manager^\conId, \ledger.\latestBock)$, discard.\\
			Otherwise, set $\manager^\conId.\conPool \define \emptyset$.
		};
		\node[event-enter] at (4, -6.8){};
		
		\end{tikzpicture}
	\end{center}
	\caption{Detailed creator challenge protocol.}
	\label{fig:diagram-creation-challenge}
\end{figure}
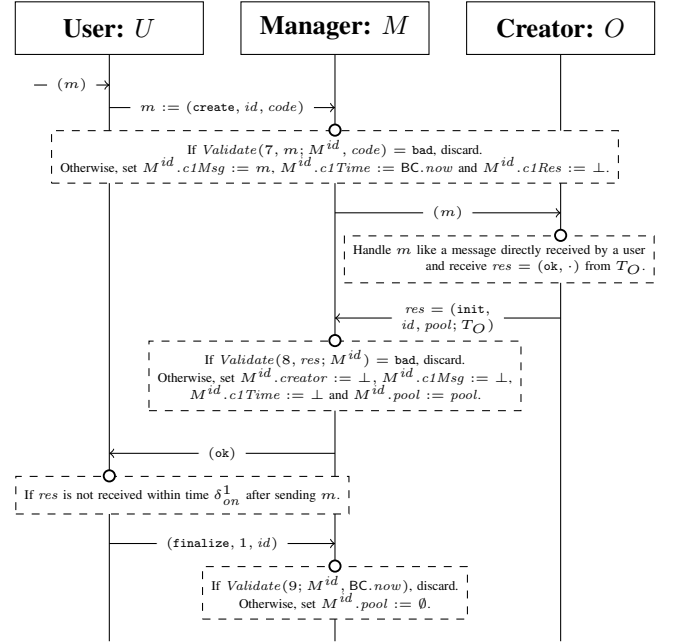

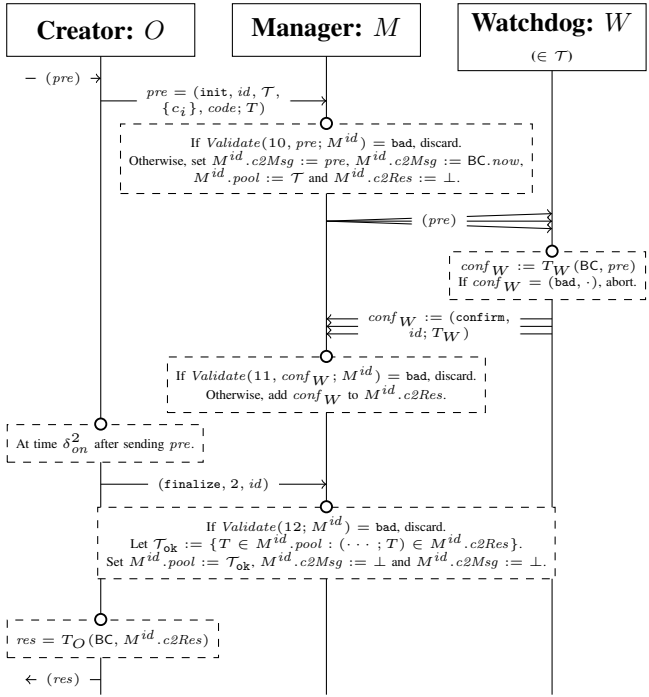
\begin{figure}
	\begin{center}
		\begin{tikzpicture}
		\newcommand{\rightMost}{8.25};
		\newcommand{\leftMost}{-0.25};
		
		\tikzset{header/.style={draw, font=\bf, fill = white, below, align=center, minimum width=2.5cm,minimum height=0.7cm}}
		\tikzset{event/.style={draw, font=\tiny, fill=white, dashed, below, align=center, minimum width=2cm,minimum height=0.5cm}}
\tikzset{arrowLabel/.style={draw, draw = white, align=center, fill=white,font=\tiny}}
\tikzset{event-right/.style={draw, font=\tiny, fill=white, dashed, below left, align=right, minimum width=2cm,minimum height=0.5cm}}
\tikzset{event-left/.style={draw, font=\tiny, fill=white, dashed, below right, align=left, minimum width=2cm,minimum height=0.5cm}}
\tikzset{event-enter/.style={draw, circle, line width=0.25mm, fill=white,inner sep=1.5pt}}

		\draw (1,-8.7) -- (1,0);
		\draw (4,-8.7) -- (4,0);
		\draw (7,-8.7) -- (7,0);
		
		\node[header] at (1,0.5) {Creator: $O$};
		\node[header] at (4,0.5) {Manager: $\manager$};
		\node[header] at (7,0.5) {Watchdog: $W$ \\ \tiny{($\in \pool$)}};
		
		\draw[->] (0,-0.5) -- node[arrowLabel] {$(\pre)$} (1,-0.5);
		\draw[->] (1,-0.8) -- node[arrowLabel] {$\pre = (\msgInit, \conId, \pool,$\\$\{c_i\}, \code; \tee)$} (4,-0.8);
		
		\node[event] at (4,-1.1) {If $\Validate(10, \pre; \manager^{\conId}) = \msgBad$, discard.\\
			Otherwise, set $\manager^{\conId}.\conWatchChalMsg \define \pre$, $\manager^{\conId}.\conWatchChalMsg \define \ledger.\latestBock$,\\
			$\manager^{\conId}.\conPool \define \pool$ and $\manager^{\conId}.\conWatchChalRes \define \bot$.
		};
		\node[event-enter] at (4, -1.1){};
		
		\draw[->] (4,-2.4) -- (7,-2.3);
		\draw[->] (4,-2.4) -- (7,-2.5);
		\draw[->] (4,-2.4) -- node[arrowLabel] {$(\pre)$} (7,-2.4);
		
		\node[event-right] at (\rightMost,-2.8) {$\conf_W \define \tee_W(\ledger, \pre)$\\
			If $\txt{conf}_W = (\msgBad, \cdot)$, abort.};
		\node[event-enter] at (7, -2.8){};
		
		\draw[->] (7,-3.7) -- (4,-3.7);
		\draw[->] (7,-3.9) -- (4,-3.9);
		\draw[->] (7,-3.8) -- node[arrowLabel] {$\conf_W \define (\msgConfirm,$\\$\conId; \tee_W)$} (4,-3.8);
		
		\node[event] at (4,-4.2) {If $\Validate(11, \conf_W; \manager^{\conId}) = \msgBad$, discard.\\
			Otherwise, add $\conf_W$ to $\manager^{\conId}.\conWatchChalRes$.
		};
		\node[event-enter] at (4, -4.2){};
		
		\node[event-left] at (\leftMost,-5.1) {At time $\timeOnchainPropagation$ after sending $\pre$.};
		\node[event-enter] at (1, -5.1){};
		
		\draw[->] (1,-5.9) -- node[arrowLabel] {$(\msgFinalize, 2, \conId)$} (4,-5.9);
		
		\node[event] at (4,-6.2) {
			If $\Validate(12; \manager^{\conId}) = \msgBad$, discard.\\
			Let $\teeSet_\msgOk \define \{\tee \in \manager^{\conId}.\conPool: (\cdots;\tee) \in \manager^{\conId}.\conWatchChalRes\}$.\\
			Set $\manager^{\conId}.\conPool \define \teeSet_\msgOk$, $\manager^{\conId}.\conWatchChalMsg \define \bot$ and $\manager^{\conId}.\conWatchChalMsg \define \bot$.
		};
		\node[event-enter] at (4, -6.2){};

		\node[event-left] at (\leftMost,-7.7) {$\res = \tee_O(\ledger, \manager^{\conId}.\conWatchChalRes)$};
		\node[event-enter] at (1, -7.7){};
		
		\draw[<-] (0,-8.5) -- node[arrowLabel] {$(\res)$} (1,-8.5);
		\end{tikzpicture}
	\end{center}
	\caption{Detailed pool challenge protocol.}
	\label{fig:diagram-creation-watchdog-challenge}
\end{figure}

In the following, we will describe the creation protocol and the corresponding challenges in detail.
The Figures~\ref{fig:diagram-creation}, \ref{fig:diagram-creation-challenge}, and \ref{fig:diagram-creation-watchdog-challenge} specify the creation protocol, the challenge of the creator and the challenge of the pool members during creation, respectively.
Program \ref{prg:program specification 2} specifies the parts of the \sysname program that are relevant for the creation.

The creation is initiated by a user $\user$ that wants to install a contract with program code $\code$.
The initial state of the contract is hard-coded in $\code$.
$\user$ obtains the set of registered enclaves from the manager $\manager$.
$\user$ is free to choose one of these enclaves as creator $\enclave_C$.
Next, a creation initialization message is sent from the user to the manager $\manager$.
This message contains a hash of $\code$ as well as the selected creator enclave.
$\manager$ picks an unused contract id $\conId$ and initializes the on-chain information about the contract including the information received from $\user$ and the latest block number $p$.
The manager returns $\conId$ and $p$ to the user.

Next, user $\user$ sends a creation request containing the contract id $\conId$ and the program code $\code$ to
the creator enclave
$\enclave_C$.
Upon receiving a creation request, $\enclave_C$ randomly selects $\poolsize$ enclaves which will form the smart contract execution pool.
One of pool enclaves is assigned as the executor enclave for contract $\conId$ while all others act as watchdog enclaves.
In addition, $\enclave_C$ samples a symmetric pool encryption key.
The generated information, i.e., the execution pool, the assigned executor enclave, and the encryption key, as well as the creation request is distributed to all pool enclaves.

Upon receiving the message from $\enclave_C$, each pool enclave  executes $\initContract(\code)$ which creates a new instance of the contract defined by $\code$.
The method locally allocates memory to set up the initial state of the contract.
Afterwards, a confirmation message is sent back to the creator enclave.

If $\enclave_C$ receives confirmations from all pool enclaves in a predefined time period, $\enclave_C$ finally generates a successful creation statement.
Otherwise, if any pool enclave has not responded timely, the creator enclave starts a challenge-response protocol.
The challenge procedures during the creation process are similar to the ones of the execution protocol explained in Section~\ref{sec:high level protocol/challenge-response}.
The differences to the challenge procedures are captured in the protocol specifications in Figure~\ref{fig:diagram-creation-challenge} and \ref{fig:diagram-challenge-watchdogs}.

Eventually, the creation statement is sent to the manager $\manager$ which updates the on-chain information about the contract.
In particular, $\manager$ marks the creation as final and stores the pool information.
If the creation process is not finalized by the creator enclave in time, user $\user$ starts the challenge-response protocol (see Section~\ref{sec:high level protocol/challenge-response}).
If the creator does not respond to the creation challenge, the creation has failed and $\user$ has to try again.

\subsubsection{\sysname Program (Creation)}
All enclaves participating in our \sysname protocol need to run a specific program.
We divided the program description into Program~\ref{prg:program execution} and~\ref{prg:program specification 2}.
The later one contains all methods executed during the creation protocol.
In particular, the creator enclave invokes the \sysname program with input $\msgCreate$, $\conId$ and contract code $\code$, where $\conId$ denotes the id assigned by the manager $\manager$.
After an execution pool of size $\poolsize$ and an encryption key are sampled, the \sysname program returns message $\msgInit$.
This message is forwarded to all execution pool members.
Upon receiving message $\msgInit$, the \sysname program initializes the new contract and returns a confirmation message.
When invoked with the confirmation messages, the creator enclave checks if a confirmation of each pool member has been received.
Only if this check is true, it returns a successful creation statement.

\begin{boxProgram}[float,label={prg:program specification 2}]{\sysname program (creation) executed by enclave $T$}
	\raggedright
	Upon receiving $(\msgCreate, \conId, \code)$, do:
	\begin{enumerate}
		\item If $\manager^{\conId} = \bot$, $\manager^{\conId}.\conCreator \neq \tee$ or $\manager^{\conId}.\conCodeHash = \Hash(\code)$, return $(\msgBad)$.
		\item Generate an encryption key $\key^{\conId}$ and sample a random subset $\pool^{\conId}$ of size $\poolsize$ from $\manager.\tees$.
		\item Let $\tee_i$ be the $i$-th member in $\pool^{\conId}$ and $\txt{pk}_i$ this member's public encryption key. Calculate $c_i \define \Enc(\key^{\conId}; \txt{pk}_i)$ for each $i$.
		\item Store $\pool^{\conId}$ and $\openResponses^{\conId} \define \pool^{\conId}$ and return $(\msgInit, \conId, \pool^{\conId}, \{c_i\}_i, \code; \tee)$.
	\end{enumerate}
	
	Upon receiving $(\msgInit, \conId, \pool, \{c_j\}_j, \code; \tee')$, do:
	\begin{enumerate}
		\item If $\tee \notin \pool$, $\tee' \neq \manager^{\conId}.\conCreator$ or $C^{\conId} \ne \bot$, return $(\msgBad)$.
		
		\item Let $T$ be the $i$-th element in $\{c_j\}_j$. Calculate and store $\key^{\conId} \define \Dec(c_i;\txt{sk}_T)$, $C^{\conId} \define \initContract(\code)$ and $\openResponses^{\conId} \define \emptyset$.
		
		\item Return $(\msgConfirm,\conId;\tee)$.
	\end{enumerate}
	
	Upon receiving message $\{m_i = (\flag_i,\conId; \tee_i)\}_i$, do:
	\begin{enumerate}
		\item If $\openResponses^{\conId} = \emptyset$, $\manager^{\conId}.\conCreator \neq \tee$, return $(\msgBad)$.
		
		\item Set $\openResponses^{\conId} = \openResponses^{\conId} \cap \manager^{\conId}.\conPool$.
		
		\item For each $m_i$ do:
		\begin{itemize}
			\item If $\tee_i \notin \openResponses^{\conId}$, skip $m_i$.
			\item Otherwise remove $\tee_i$ from $\openResponses^{\conId}$.
		\end{itemize}
		
		\item If $\openResponses^{\conId} \neq \emptyset$, return $(\msgBad)$.
		Otherwise, return $(\msgInit, \conId, \pool^{\conId};\tee)$.
		
		\item If $T \notin \pool^{\conId}$, delete all stored variables with respect to $\conId$.
	\end{enumerate}
\end{boxProgram}

\subsubsection{Validation}

All of the different messages sent to the manager throughout the protocol need to be validated with several checks.
In order to keep the description compact, we did not include the validation steps in the protocol figures but extracted them into a validation algorithm specified in Program~\ref{prg:verify-program}.
The algorithm is invoked with an counter specifying the checks that should be performed, an optional message that should be checked and the contract state tuple maintained by the manager.
The validation returns $\msgOk$ if all requirements are satisfied and $\manager$ can continue executing and $\msgBad$ if $\manager$ should discard the received request.

	\begin{boxProgram}[float,label={prg:verify-program}]{Algorithm $\Validate$}
		\raggedright
		The validation algorithm performs the following checks.
		If input $C = \bot$, the parsing of a message fails or any require is not satisfied, the algorithm outputs $\msgBad$.
		Otherwise, it outputs $\msgOk$.
		\begin{itemize} [leftmargin=*]
			\item On input $(1, m; C)$, parse $m$ to $(\msgExecute, \conId, \cdot, \cdot; U)$.
			Require that $C.\conCreator = \bot$, $C.\conExecChalBlock = \bot$ and $\Verify(m) = \msgOk$.  
			\item On input $(2, \res; C)$, parse $\res$ to $(\msgOk, \conId, \cdot, h; \tee)$. Require that $C.\conCreator = \bot$, $\Hash(C.\conExecChalMsg) = h$, $C.\conExecChalBlock + \timeOnchainExecution > \ledger.\latestBock$, $\Verify(\res) = \msgOk$ and $C.\conPool[0] = \tee$.  	
			\item On input $(3; C)$, require that $C \ne \bot$, $C.\conCreator = \bot$, $C\conExecChalMsg \ne \bot$ and $C.\conExecChalBlock + \timeOnchainExecution \leq \ledger.\latestBock$.  	
			
			\item On input $(4, \pre; C)$, parse $\pre$ to $(\msgUpdate,\conId,c,h; \tee)$. Require that $C.\conCreator = \bot$, $C.\conWatchChalBlock = \bot$, $C.\conPool[0] = \tee$ and $\Verify(\pre) = \msgOk$.  
			\item On input $(5, \conf; C)$, parse $\conf$ to $(\msgConfirm,\conId, h; \tee_i)$ and $C.\conWatchChalMsg$ to $(\cdot, \cdot, \cdot, h'; \cdot)$.
			Require that $C.\conCreator = \bot$, $C.\conWatchChalBlock + \timeOnchainPropagation > \ledger.\latestBock$, $\Verify(\conf) = \msgOk$, $h = h'$ and $\tee \in C.\conPool$.  
			\item On input $(6; C)$, require that $C.\conCreator = \bot$, $C.\conWatchChalBlock \ne \bot$ and $C.\conWatchChalBlock + \timeOnchainPropagation \leq \ledger.\latestBock$.  
			
			\item On input $(7, m; C, \code)$, parse $m$ to $(\msgCreate, \conId, \code)$.
			Require that $C.\conCreator \ne \bot$, $C.\conExecChalBlock = \bot$ and $C.\conCodeHash = \Hash(\code)$.  
			\item On input $(8, \res; C)$, parse $\res$ to $(\msgInit, \conId, \pool; \tee)$. Require that $C.\conCreator = \tee$, $C.\conExecChalBlock + \timeOnchainCreation > \ledger.\latestBock$ and $\Verify(\res) = \msgOk$. 	
			\item On input $(9; C)$, require that $C.\conCreator \ne \bot$, $C\conExecChalBlock \ne \bot$ and $C.\conExecChalBlock + \timeOnchainCreation \leq \ledger.\latestBock$.  
			
			\item On input $(10, \pre; C)$, parse $\pre$ to $(\msgInit, \conId, \cdot, \cdot, \cdot; \tee)$. Require that $C.\conCreator = \tee$, $C.\conWatchChalBlock = \bot$ and $\Verify(\pre) = \msgOk$ .  
			\item On input $(11, \conf; C)$, parse $\conf$ to $(\msgConfirm,\conId; \tee_i)$. Require that $C.\conCreator \ne \bot$, $C.\conWatchChalBlock \ne \bot$, $C.\conWatchChalBlock + \timeOnchainCreationPropagation > \ledger.\latestBock$, $\Verify(\conf) = \msgOk$ and $\tee \in C.\conPool$.  	
			\item On input $(12; C)$, require that $C.\conCreator \ne \bot$, $C.\conWatchChalBlock \ne \bot$ and $C.\conWatchChalBlock + \timeOnchainCreationPropagation \leq \ledger.\latestBock$. 
		
		\end{itemize}
	\end{boxProgram}

\subsection{Timeouts}
\label{sec:high level protocol/timeouts}

Our protocol incorporates several timeouts $\delta^*_\txt{off}$, which define until when an honest user or operator expects a response to a request, and $\delta^*_\txt{on}$, which define until when the manager expects a response to a challenge.
These timeouts have to be selected carefully such that each honest party has the chance to answer each message and challenge before the respective timeout expires.
In this section, we elaborate on the requirements on the timeouts.
We neglect message transmission delays and also assume that each challenge sent to the manager will directly be received by all operators (already before it is included into a final block)\footnote{Instead, we could also add two times the maximum message delay to each off-chain timeout $\delta^*_\txt{off}$ and the blockchain confirmation time $\confTime = \blockTime \cdot \nbConfBlocks$ to each on-chain timeout $\delta^*_\txt{on}$.}.
We recall the maximum blockchain delay which is defined as $\delta_{\ledger} = \alpha \cdot \tau$ (cf.~\ref{sec:adversary-model} and \ref{sec:protocol/blockchain}).

The off-chain propagation timeout $\timeOffchainPropagation$ describes the time an execution or creation operator maximally waits for a confirmation from the (other) pool members.
It needs to be larger than the maximal update respectively installation time of a contract.
Timeout $\timeOnchainPropagation \geq \timeOffchainPropagation + \delta_{\ledger}$ describes the maximal time after which $\manager$ expects a response to any watchdog challenge, either during creation or execution.
The off-chain execution timeout $\timeOffchainExecution$ describes the maximal time a user waits for a response to an execution request.
Note that there might be a running execution and both running and new execution might require a watchdog challenge.
In case watchdogs are dropped in the process of such a challenge, the executor needs to be able to notify its enclave about the new pool constellation, and hence, wait until the finalization of the challenge is within a final block.
This takes additional time $\confTime = \blockTime \cdot \nbConfBlocks$ (cf. \ref{sec:protocol/blockchain}).
Hence, $\timeOffchainExecution$ needs to be high enough to enable the challenged executor to perform two contract executions and run two watchdog challenges each taking up to time $\timeOnchainPropagation + \delta_{\ledger} + \confTime$.
We elaborate on maximal execution, update, and installation times of contracts in Section~\ref{sec:protocol/security-remarks}.
Finally, $\timeOnchainExecution \geq \timeOffchainExecution + \delta_{\ledger}$ defines the maximal time after which $\manager$ expects a response to an execution challenge.
As the creation is comparable to the execution, we set the timeouts for off-chain creation and creation-challenge according to the ones of the execution.

	The timeouts are an upper bound of the delay that can be enforced by malicious operators by withholding messages.
	To decrease the delays in a practical setting our implementation incorporates dynamic timeouts.
	Such a timeout is initially set to match an optimistic scenario where all operators directly answer.
	Only if the executor signals that a watchdog is not responding, the timeout is increased.
	For example, the $\timeOnchainExecution$ timeout is initially set by the manager just high enough to allow the executor to perform the execution offline and to send one on-chain transaction.
	This on-chain transaction is either the response or a watchdog challenge.
	In case the executor creates a watchdog challenge this triggers the manager to increase the $\timeOnchainExecution$ timeout for the executor.
	Similarly, the timeout $\timeOnchainExecution$ is increased by the manager if any watchdog is not responding and the executor sends a transaction that kicks this watchdog.
	The increased timeout allows the executor to provide the kick transaction together with enough confirmation blocks to its enclave to finalize the execution.
	This dynamic timeout mechanism still allows the executor to respond in time even if a watchdog is not responding but at the same prevents the executor to stall the execution to the maximum although the watchdogs have already responded.
	While the executor still can create a watchdog challenge to increase the delay, this attack is costly since the executor needs to pay for the on-chain transaction.
	The value of the off-chain timeout $\timeOffchainExecution$ is handled similarly.
	The client only needs to account for watchdog challenges in the previous execution in the timeout if there is indeed a running on-chain challenge.
	If there are no running challenges, a client can decrease $\timeOffchainExecution$ to $\delta_{\ledger}$ plus two times the time it takes a TEE to execute and update a contract.
	Hence, if the executor is unresponsive, the client submits its executor challenge much earlier.

	We give a concrete evaluation for the case of Ethereum, as this is the platform on which our implementation works.
	Let $\alpha = 20$ be the number of blocks until a transaction is included in the blockchain in the worst case, and $\alpha_{\txt{avg}} = 10$ in the average case. 
	Further, we consider the block creation time to be $\tau = 44 s$ per block in the worst case and $\tau_{\txt{avg}} = 15 s$ in the average case\footnote{For setting $\alpha$ and $\alpha_{\txt{avg}}$, we consider a transaction to be included into the blockchain after at most 20 resp. 10 blocks according to \cite{blocksToMined}. To determine $\tau$, we analyzed the Ethereum history via Google-BigQuery and identified that since 2018 every interval of 20 blocks took at most $44 s$ per block. For $\tau_{\txt{avg}}$, we take the average parameter for Ethereum (cf.~\url{https://etherscan.io/chart/blocktime}).}.
	Finally, we assume that blocks are final, when they are confirmed by $\gamma =15$ successive blocks.
	Since the network delay and the computation time of enclaves are at most just a few seconds, which is insignificant compared to the time it requires to post on-chain transactions, we neglect these numbers for simplicity in the following example.
	In case the executor (resp. a watchdog) is not responding, it is challenged by the the client (resp. the executor).
	The creation of such a challenge takes $\alpha_{\txt{avg}} \cdot \tau_{\txt{avg}} = 150 s$ on average.
	In what follows, due to the dynamic timeout mechanism, the on-chain timeout for both, executor challenge ($\timeOnchainExecution$) and watchdog challenge ($\timeOnchainPropagation$), is initially set to $\alpha \cdot \tau = 880 s $.
	For on-chain timeouts, we need to consider the worst-case parameters to allow honest operators to respond timely in every situation.
	While a dishonest operator can delay up to the defined timeout, an honest operator responds, and hence, finalizes the challenge in $150 s$ on average.
	In case the challenged operator gets kicked, the (next) executor enclave needs to provide the kick transaction together with enough confirmation blocks to its enclave to finalize the execution.
	This takes $(\alpha_{\txt{avg}} + \gamma) \cdot \tau_{\txt{avg}} = 375 s$ on average.
	For executor challenges, it can happen that the executor submits a watchdog challenge during the timeout period.
	In this case, which can happen at most twice, the timeout is increased by $880 s$.
	If the challenged watchdog does not reply, and consequently is kicked from the pool, the timeout is increased by $(\alpha + \gamma) \cdot \tau = 1\,540 s$.
	Note, this worst case is very costly to provoke, and in the general case, an honest executor can finalize the kick of the watchdog in $375 s$.%

\subsection{Coin Flow}
\label{sec:protocol/payments}

The \sysname protocol supports the off-chain execution of smart contracts that deal with coins, e.g., games with monetary stakes.
To this end, we provide means to send coins to and receive coins from a contract.
In this section, we explain the mechanisms that enable the transfer of money and the intended coin flow of \sysname contracts.

In order to deposit money to a \sysname contract, identified by $\conId$, a user $\user$ sends a message $(\msgDeposit, \conId, \coins; \user)$ with $\coins$ coins to $\manager$.
Upon receiving a deposit message, $\manager$ checks whether a contract with identifier $\conId$ exists and validates the signature, i.e., $\manager^\conId \neq \bot$ and $\Verify(\msgDeposit, \conId, \coins; \user) = \msgOk$.
If the checks hold, $\manager$ increases the contract balance $\manager^\conId.\conBalance$ by $\coins$.
As deposits are part of blockchain data that are provided by the operator to an enclave (cf.~\ref{sec:high level protocol/synchronization}) and the enclave forwards the data to the $\conNextState$ function of the contract $C^\conId$, $\user$ is ensure that $C^\conId$ processes the deposit once the corresponding block is final.
However, it is upon to the application logic to decide how deposits are processed.

A contract $C$ can transfer coins to users by outputting \textit{withdrawals} as part of the public state.
It is upon the application logic to decide how and when coins are transferred to the users.
For example, a game can issue withdrawals once the winner has been determined or leave the coins locked for another round unless a user explicitly requests a withdrawal via a contract execution.
However, once a withdrawal has been issued, the coins are irreversible transferred. 

Technically, contract $C$ with identifier $\conId$ maintains a list of all unspent withdrawals $\{\coins_i, U_i\}$ and a counter $\conPayoutLevel$ for the number of spent payouts.
Each public state returned by $C$ contains a payout, a signed message $m \define (\msgWithdraw, \conId, \conPayoutLevel, \{\coins_i, U_i\}; \enclave_E)$ where $\enclave_E$ is the executor enclave of the contract.
This message can be sent to $\manager$ to spent all withdrawals within the payout.
$\manager$ checks the validity of the payout, i.e., $\Verify(m) = \msgOk$, $\enclave_E = \manager^{\conId}.\conPool[0]$, and $\conPayoutLevel = \manager^\conId.\conPayoutLevel$.
If the checks hold, $\manager$ transfers coins to the users according to the withdrawal list $\{\coins_i, U_i\}$.
Finally, $\manager$ sets $\manager^\conId.\conPayoutLevel \define \conPayoutLevel + 1$ and $\manager^\conId.\conBalance \define \manager^\conId.\conBalance - \txt{sum}$, where $\txt{sum}$ is the sum of all withdrawals.
Once $C$ processes a final block with a payout transaction, it updates its list of unspent withdrawals $\{\coins_i, U_i\}$ accordingly and increments $\conPayoutLevel$ by $1$.

This mechanism ensures that a malicious user can neither double spent withdrawals nor prevent an honest user from withdrawing his coins as long as the contract remains live.
Note that for each value of $\conPayoutLevel$ only one payout can be submitted successfully.
Further, a contract only issues a payout for the next value of $\conPayoutLevel$ once it has processed a final block containing the current value of $\conPayoutLevel$.
As the contract removes the already spent withdrawals from the list, it prevents that any withdrawal is double spent.
Although a payout temporarily invalidates all other payouts for the same value of $\conPayoutLevel$ and hence might invalidate same withdrawals, the unspent withdrawals will be included in each payout of the incremented $\conPayoutLevel$ and hence are spent with the next payout submission.

\begin{figure}[t]
	\centering
	\includegraphics[width=0.85\columnwidth, trim=0.2cm 0.3cm 0.3cm 0.2cm, clip]{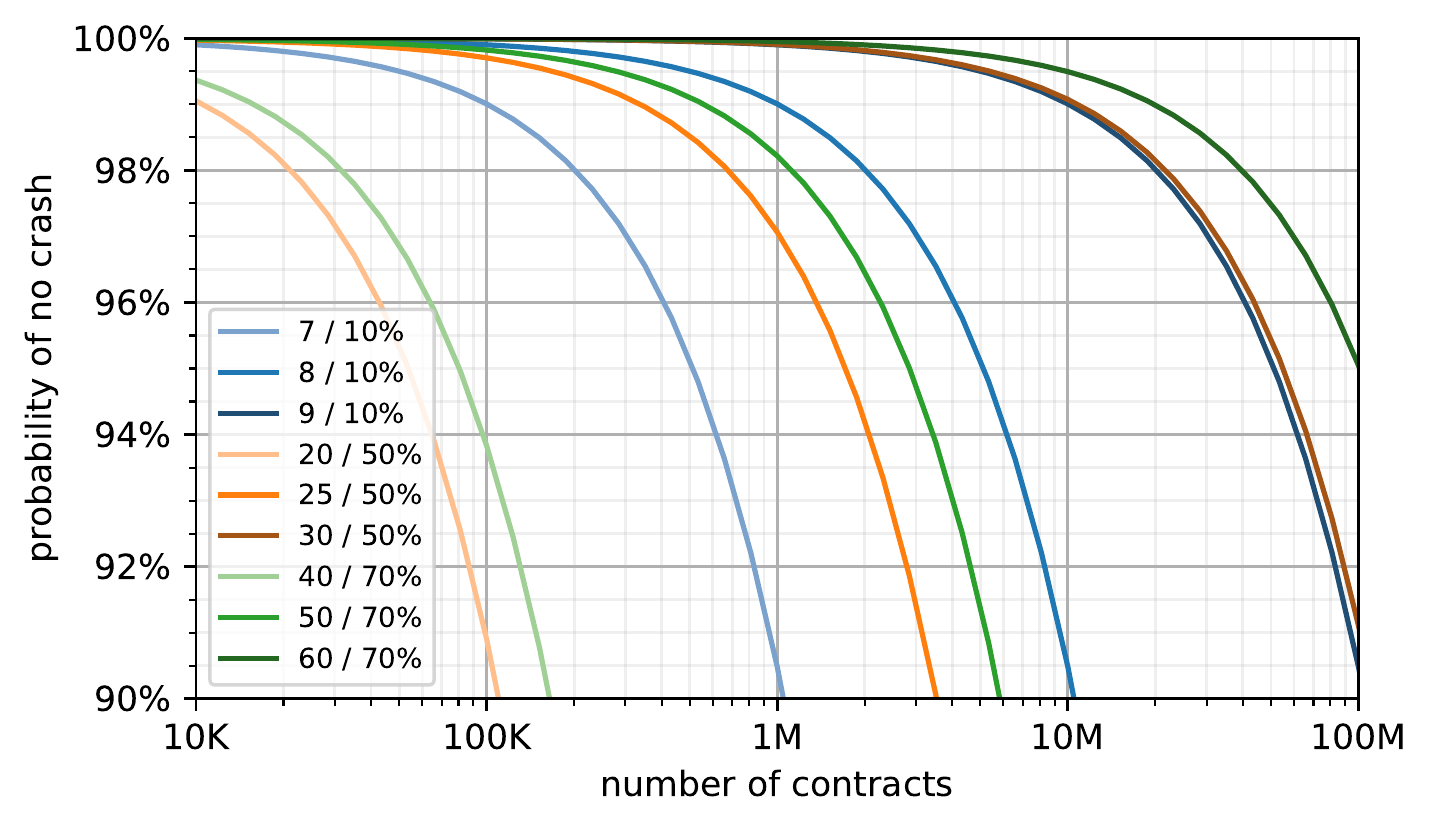}
	\caption{Cumulative probabilities of no contracts crashing with a large number of POSE contracts in the system for different pool sizes $s$ and adversary shares $m$, labeled ``$s$ / $m$''.}
	\label{fig:probs}
\end{figure}

\end{document}